\def\apjs{ApJS}                                     % ApJS
\def\aap{AAP}                                       % AAP
\def\lsim{\lower.5ex\hbox{$\; \buildrel < \over \sim \;$}}
\def\gsim{\lower.5ex\hbox{$\; \buildrel > \over \sim \;$}}
\documentclass{aa}
\usepackage{epsfig}

\begin{document}

%\thesaurus{}
%%%%%%%%%%%%%%%%%%%%%%%%%%%%%%%%%%%%%%%%%%%%%%%%%%%%%%%%%%%%%%%%
% The titlepage
%%%%%%%%%%%%%%%%%%%%%%%%%%%%%%%%%%%%%%%%%%%%%%%%%%%%%%%%

\title{The lives and deaths of positrons in the interstellar medium}

\author{N.~Guessoum\inst{1}
         \and P.~Jean\inst{2}
         \and W.~Gillard\inst{2}}
\institute{ 
$^{1}$ American University of Sharjah, College of Arts \& Sciences, 
Physics Department, PO Box 26666, Sharjah, UAE \\
$^{2}$ CESR, CNRS/UPS, B.P.~4346, 31028 Toulouse Cedex 4, France \\
}

   \date{Received  ; accepted }

%   \authorrunning{N.~Guessoum, P. Jean \&\ W. Gillard}
   \authorrunning{Guessoum, Jean \&\ Gillard}

   \titlerunning{The lives and deaths of positrons in the ISM}

%%%%%%%%%%%%%%%%%%%%%%%%%%%%%%%%%%%%%%%%%%%%%%%%%%%%%%%%%%%%%%%%%%%%%%%%%%
% Abstract
%%%%%%%%%%%%%%%%%%%%%%%%%%%%%%%%%%%%%%%%%%%%%%%%%%%%%%%%%%%%%%%%%%%%%%%%%%

\abstract{   

We reexamine in detail the various processes undergone by positrons 
in the interstellar medium (ISM) from their birth to their 
annihilation using the most recent results of positron interaction cross 
sections with atomic and molecular hydrogen, as well as helium. 
The positrons' lives are divided into two phases: the ``in-flight" phase 
(between $\approx 1$ MeV and tens of eV) and the thermal phase. 
The first phase is treated with a Monte Carlo simulation that 
allows us to determine the fraction of positrons that form positronium 
and annihilate as well as the characteristics of the annihilation 
emission as a function of the medium conditions. The second phase is 
treated with a binary reaction rate approach, with cross sections adopted 
from experimental measurement or theoretical calculations. An extensive 
search and update of the knowledge of positron processes was thus 
undertaken. New reaction rates and line widths have been obtained.

We investigate the treatment of the 
complicated interactions between positrons and interstellar dust grains. 
Fully relevant data were not always available, but we were nonetheless 
able to reach satisfactory understanding of positron annihilation
on grains, both qualitatively 
and quantitatively. All factors of the problem have been considered, 
including the grain size distribution and composition, the electric charge 
of the grains, the backscattering, positronium formation and ejection from 
the grain, the pick-off annihilation inside them and the partial 
destruction of dust in the hot regions of the ISM. New reaction rates 
and widths of the line resulting from the annihilation inside and outside 
of the grain have been obtained. The final results of our calculations 
(reaction rates and spectra) showed that dust is only important in the 
hot phase of the ISM, where it dominates all other processes.

Combining the new calculations, we have constructed annihilation 
spectra for each phase of the ISM, considering various grain contents, 
as well as an overall combined spectrum for the ISM as a whole.

\keywords{Gamma rays: theory -- Line: formation, profile -- ISM: general}

}

\maketitle

%%%%%%%%%%%%%%%%%%%%%%%%%%%%%%%%%%%%%%%%%%%%%%%%%%%%%%%%%%%%%%%%%%%%%%%%%%
% 1. Introduction
%%%%%%%%%%%%%%%%%%%%%%%%%%%%%%%%%%%%%%%%%%%%%%%%%%%%%%%%%%%%%%%%%%%%%%%%%%

\section{\label{s1}Introduction}

Positrons are produced at tremendous rates in the Galaxy. This has been 
firmly established by a long series of balloon and satellite detections of 
the positrons' landmark signature, the gamma-ray line at 511 keV produced 
in their annihilation with electrons. A rate of $\approx 10^{43} e^+/s$ is 
usually inferred from such detections (Purcell et al. 1997, Milne et al. 
2000). 
Surprisingly, despite three 
decades of (indirect) detections of positrons, the origin and galactic 
distribution of these huge amounts of antiparticles is still puzzling 
astrophysicists. There are several major reasons for this problem: 1) our 
understanding of the positron production processes in astrophysical media 
is plagued by significant uncertainties; 2) the galactic map of the 511 keV 
annihilation line, which is supposed to lead us to the positrons, has yet 
to be determined with any reasonable accuracy; 3) the physics of the 
annihilation processes, which have until now been believed to be well 
understood, may need to be improved to a certain extent; 4) our lack of 
knowledge of the extent of diffusion/propagation of positrons between 
their production sites and their annihilation places.  

In nature, positrons can be produced by one of the following main 
processes: a) $\beta^+$ decay of radioactive nuclei; b) $\pi^+$ decay into 
$\mu^+$, which decays and gives off a positron; c) pair 
(electron-positron) production through photon-photon interactions; d) pair 
production by the interaction of an electron with a strong magnetic field. 
All these processes can be found in astrophysical settings, provided that 
the physical conditions for them are fulfilled. For example, the first 
process (the $\beta^+$ decay of a radioactive nucleus) is widespread in 
the 
explosive and/or hydrostatic nucleosynthesis environments of novae, 
supernovae, Wolf-Rayet and Asymptotic Giant Branch stars. The second 
process ($\pi^+ \rightarrow \mu^+ \rightarrow e^+$) takes place where the 
pion is produced in collisions of highly energetic (more than $\approx 
200$ MeV) 
cosmic rays with interstellar material (mostly protons). The third process 
(photon-photon pair production) requires photons of high energies and is 
known to occur in the environment of luminous compact objects (black hole
candidates, micro-quasars, active galactic nuclei, etc.).  The fourth 
process requires intense magnetic fields and is rather common in the 
environments of pulsars/neutron stars.

On the scale of the whole Galaxy, as we are interested in here, the main 
positron generators are, in order of decreasing importance: supernovae 
(SNe), novae (Ne), compact objects (microquasars, pulsars, etc.), cosmic 
rays and possibly gamma ray bursts. However, substantial uncertainties 
over the specific factors involved in each case have made the precise 
identification of the positron sources very difficult. For instance, for 
the main candidates for galactic positrons (supernovae), depending 
on their type, the specific positron production process and the escape 
fraction from the expanding shell into the interstellar medium (ISM), 
the amount of positrons contributed by supernovae may vary significantly.

The issue of the origin of positrons responsible for the observed 
annihilation in the Galaxy will be investigated in another work, in 
relation to the new 511 keV map(s) obtained by INTEGRAL/SPI (Knodlseder 
et al. 2005). For a recent review of some of the issues pertaining to this 
topic, see the paper by Paul (2004) or Jean et al. (2004). 
We will address here only the physics 
of the annihilation as it leads to the spectra observed by high 
energy-resolution instruments.

Since the discovery of the 511 keV annihilation radiation emanating 
from the Galactic center in 1970 (Johnson, Harnden and Haymes 1972), 
interest in this radiation, from the observational viewpoint as well as 
from the theoretical stance, has been very high. 

On the observational side, dozens of balloon and, more recently, 
satellite missions have been conducted over the past 2 or 3 decades. We 
will cite only high energy-resolution (germanium) instruments that allowed 
for the extraction of a detailed line spectrum, because 
the physics of the annihilation produces a spectrum that can only be 
compared to one obtained by such instruments. In 1977 the Ge-detector 
(high energy resolution) balloon mission 
(Leventhal, MacCallum and Stang 1978), established the electron-positron 
annihilation nature of the radiation by accurately measuring the line 
centroid energy at 511 keV. The Gamma-Ray Imaging Spectrometer (GRIS) 
balloon missions (Gehrels et al., 1991; Leventhal 
et al. 1993) firmly established the narrow width of the line (FWHM less 
than about 3 keV) and separated the extended, diffuse emission in the 
interstellar medium (ISM) from that of the Galactic center region. The 
HEXAGONE balloon-borne spectrometer mission (Chapuis et al., 1991; 
Durouchoux et al. 1993) also detected a narrow line (FWHM less than about 
3 keV). The Transient Gamma-Ray Spectrometer (TGRS), onboard the WIND 
satellite, rather strongly disproved the two-component, time-variability (of 
the Galactic Center emission) paradigm (Teegarden et al., 1996; Harris et 
al. 1998; Harris 2000) and showed an even narrower line (FWHM $\approx 1.8$ 
keV). The most recent analysis of the data from the on-going 
INTEGRAL/SPI detection of the annihilation line shows the line to have a 
width of $\approx$ 2.3 keV (Knodlseder et al. 2005).

In parallel, theoretical modeling and data analyses of the emission have 
proceeded in an attempt to understand both the physical 
processes behind the emission and the physical conditions prevailing in 
the regions where the annihilation is taking place. 

Detailed theoretical investigations of the annihilation processes behind 
the observed radiation go back to Crannell et al. (1976) who, while 
interested in the line produced specifically in solar flares, investigated 
most of the processes involved in electron-positron interactions. Other 
researchers, i.e. Steigman (1968), 
Stecker (1969) and Leventhal (1973) had pointed out the importance of the 
Positronium in astrophysical settings. Positronium (Ps) is the bound state 
of a positron with an electron, and depending on whether the two have 
parallel or anti-parallel spins, this ``atom" will be in the long-lifetime 
(t = 140 ns) ortho-Ps state or the short-lifetime (t = 0.125 ns) para-Ps 
state. The latter decays into 2 photons of 511 keV (plus or minus some 
small amount of energy corresponding to the relative momenta of the bound 
particles), while the former (ortho-Ps) state decays into 3 photons of 
energies totaling 1022 keV. The distribution of photon energies resulting 
from the decay of ortho-Ps was first calculated by Ore \& Powell (1949) 
and more recently by Adkins (1983).

Subsequent work on the positron processes focused on the conditions of the 
interstellar medium. Bussard, Ramaty and Drachman (1979 -- hereafter 
BRD79) 
calculated the rates of the various processes undergone by low-energy 
positrons and obtained the resulting annihilation spectrum; Ramaty \& 
Meszaros (1981) considered the processes and the resulting radiation 
in the conditions of a relativistic, hot (high-temperature) plasma. 
Zurek (1985) emphasized the 
role of dust grains and showed their potentially important modification of 
the annihilation spectrum, especially in the hot/warm/ionized phases of 
the ISM. Guessoum, Ramaty and Lingenfelter (1991) reconsidered the 
problem in light of earlier developments (new calculations of reaction 
rates by Gould 1989, recent observational data, more precise treatment of 
dust grains, etc.), performed detailed theoretical spectroscopy of the 
annihilation line, and produced spectra for various conditions of the ISM; 
one of the ideas emphasized in that work was that the FWHM value was not 
an accurate descriptor of the line, as the line cannot always be simply 
represented or fitted by a Gaussian function. Guessoum, Ramaty and Skibo 
(1997) updated the knowledge of positron processes (reaction rates and 
line widths), adding molecular hydrogen and helium. Most recently, Dermer 
\& Murphy (2001) reviewed the positron production and annihilation 
mechanisms, dealing with both the sources of positrons and the line 
profile depending on the annihilation environment.

Positron annihilation models for astrophysical settings also rely heavily 
on many laboratory measurements and theoretical calculations of positron 
interactions with atoms and molecules. We mention the pioneering work of: 
Bhatia, Drachman and Temkin (1977) for early theoretical 
calculations of annihilation in positron-hydrogen collisions; Brown et al. 
(1984), Brown, Leventhal and Mills (1986), Brown \& Leventhal (1986) for 
measurements of line widths and Positronium formation fractions in gases 
(H$_2$, He) relevant to the ISM conditions; see Charlton \& Humberston 
(2001) for a general review of positron work in gases (particularly H, 
H$_2$, He), Puska \& Nieminen (1994) for the theory of positrons in solids 
and on solid surfaces, and the body of work by Iwata, Surko and 
collaborators (Iwata et al. 1996, Iwata et al. 1997, Iwata et al. 2000) 
on laboratory measurements of cross sections and annihilation line widths 
of positrons colliding with atoms and molecules (up to and including 
large 
complex alcohol and/or aromatic molecules, some of which may exist in 
galactic molecular clouds). We would draw attention to the significant 
amount of  relevant ``non-astrophysical" work currently being 
performed by atomic and condensed 
matter researchers in positron physics \footnote{See the work by the 
groups at 
University of California, San Diego 
(http://physics.ucsd.edu/research/surkogroup/positron/), Surko et al. 
(2000); the groups at Martin-Luther-University Halle 
(http://positron.physik.uni-halle.de/), at University of Michigan 
(http://positrons.physics.lsa.umich.edu/); and others.}. 

In order to determine the overall 
spectrum of the line produced by the galactic positrons, one must consider 
a model of the ISM, i.e. distributions and characteristics (temperature, 
ionization, etc.) of the gas (H, H$_2$, He) and dust. Guessoum, Ramaty 
and Lingenfelter (1991) based their work on the model of McKee \& Ostriker 
(1977); Gillard (pers. comm.) use the models of Ferriere 
(1998a, 1998b, 2001), deeming them more detailed and more conveniently 
described by analytical expressions.

Taking all the above physical information (processes, cross sections, line 
widths, gas distributions, etc.) into account, one can attempt to 
reproduce the observational data (from GRIS, HEXAGONE, TGRS, INTEGRAL/SPI) 
and infer useful knowledge about the environment where the annihilation is 
predominantly taking place. This kind of approach was adopted by Wallyn et 
al. (1993) for the HEXAGONE data and by Guessoum et al. (2004) for the SPI 
and the TGRS data.

The latest such analysis (Guessoum et al. 2004) 
partly led us to revisit the subject of positron processes and annihilation. 
Analysis of the SPI data (for which the FWHM was then 
(2.67$\pm$0.33) keV - Lonjou et al. 2004), unlike that of TGRS (for which 
the FWHM was 
(1.8$\pm$0.5) keV - Harris et al. 1998) tended to imply a very large 
amount (or 
reaction rate) of dust grains, simply because the interaction of positrons 
with grains resulted (in the treatments done until now) in a line made up 
of two Gaussians, one with $\Gamma = 2.5 $ keV and one with $\Gamma = 
1.8$ keV. Similarly, the cross section for positron-hydrogen charge 
exchange had been known to be extremely important, because this reaction 
dominates all other processes in the cold phases of the ISM, and because 
it leads to both a wide line (6.4 keV in the old treatments) and a large 
Ps fraction (f$_1$, the fraction of the positrons that annihilate via 
Positronium formation). We thus realized that a detailed and refined 
treatment of positron processes was needed before one can attempt to 
analyze the high-quality data and draw astrophysical conclusions from it.

In the next section, we review 
the processes that take place ``in flight", i.e. before the positrons 
thermalize with the medium of the given ISM phase; this applies for 
positron energies between $\approx$ 1 MeV (their kinetic energies when 
they come out of beta decay) and a few tens of eV, when Positronium 
formation through 
charge exchange starts to occur. This regime includes Coulomb energy 
losses (with free electrons), excitation and ionization of atoms (H, 
H$_2$, He), and Ps formation by charge exchange with H, H$_2$ and He. We 
review the cross sections for all the relevant processes and perform a 
Monte Carlo simulation to determine the width of the line produced in this 
regime. This is done with utmost care 
because of the importance of the charge exchange process as explained 
above. In section 3, we review the positron processes after 
thermalization; these comprise: charge exchange with H, H$_2$, He, direct 
annihilation with bound electrons, direct annihilation with free 
electrons and radiative combination (with free electrons). We discuss
the dust grains and the processes that positrons 
undergo with them. We also 
present up-to-date line widths for all processes. In section 4, we use the 
knowledge compiled and presented here to obtain annihilation spectra in 
standard ISM situations. In the final section, we summarize the most 
important new information and discuss the application of this revision to 
the 511 keV SPI data analysis that we intend to perform.

%%%%%%%%%%%%%%%%%%%%%%%%%%%%%%%%%%%%%%%%%%%%%%%%%%%%%%%%%%%%%%%%%%%%%%%%%%
% 2. In-Flight Processes
%%%%%%%%%%%%%%%%%%%%%%%%%%%%%%%%%%%%%%%%%%%%%%%%%%%%%%%%%%%%%%%%%%%%%%%%%%

\section{\label{s2}In-flight processes}

When, after being slowed down by Coulomb collisions in the ISM, 
the positron has an energy less than a few tens of eV, it starts to pick 
up 
an electron from an atom or a molecule by charge exchange, thus forming 
a positronium. This process is endoenergetic; it can happen as long as 
the kinetic energy of the positron is above the charge exchange threshold 
energy of Ps formation with the given atom or the molecule. We calculate 
the probability for such an event to occur using the Monte Carlo method, 
similarly to the method presented in BRD79. In the following sections we 
present the cross sections we used for the calculation and the results of 
the simulation for positrons in atomic and molecular hydrogen and helium.

%%%%%%%%%%%%%%%%%%%%%%%%%%%%%%%%%%%%%%%%%%%%%%%%%%%%%%%%%%%%%%%%%%%%%%%%%%
% 2.1 Energy losses and cross-sections
%%%%%%%%%%%%%%%%%%%%%%%%%%%%%%%%%%%%%%%%%%%%%%%%%%%%%%%%%%%%%%%%%%%%%%%%%%

\subsection{Energy losses and cross-sections}
\begin{figure}[tb]
%\centering
\includegraphics[width=8.8cm,height=6.0cm]{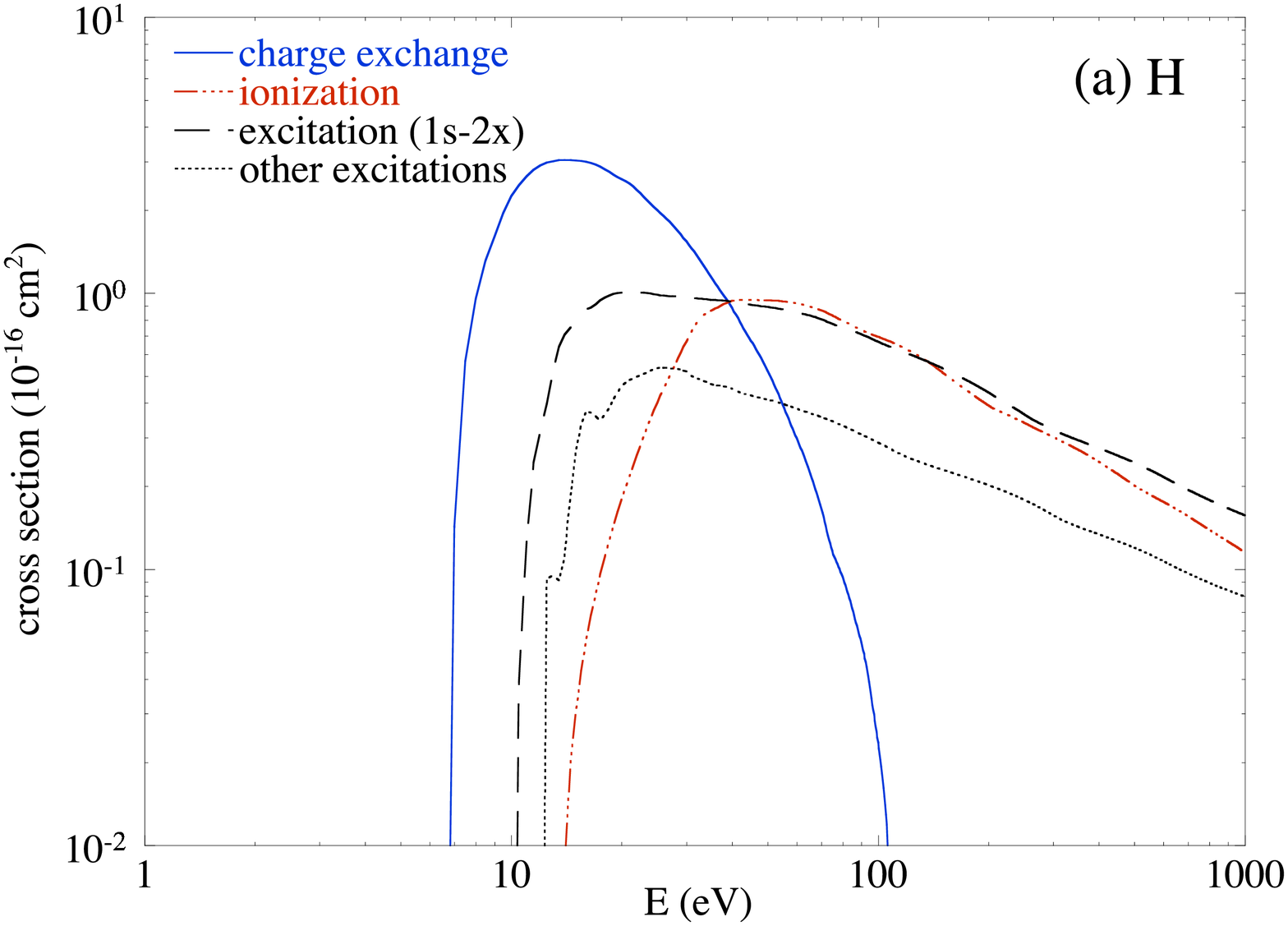}
\includegraphics[width=8.8cm,height=6.0cm]{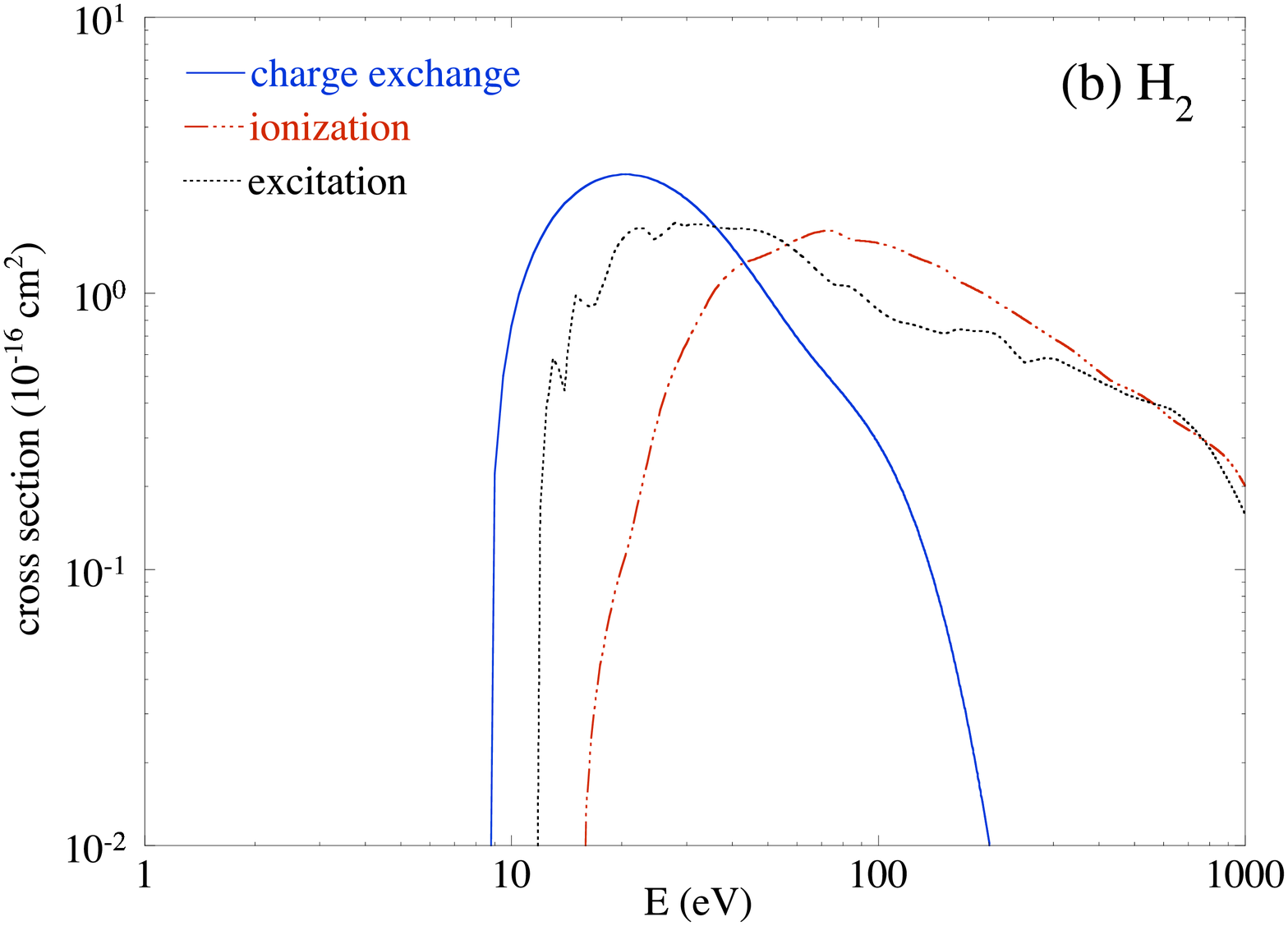}
\includegraphics[width=8.8cm,height=6.0cm]{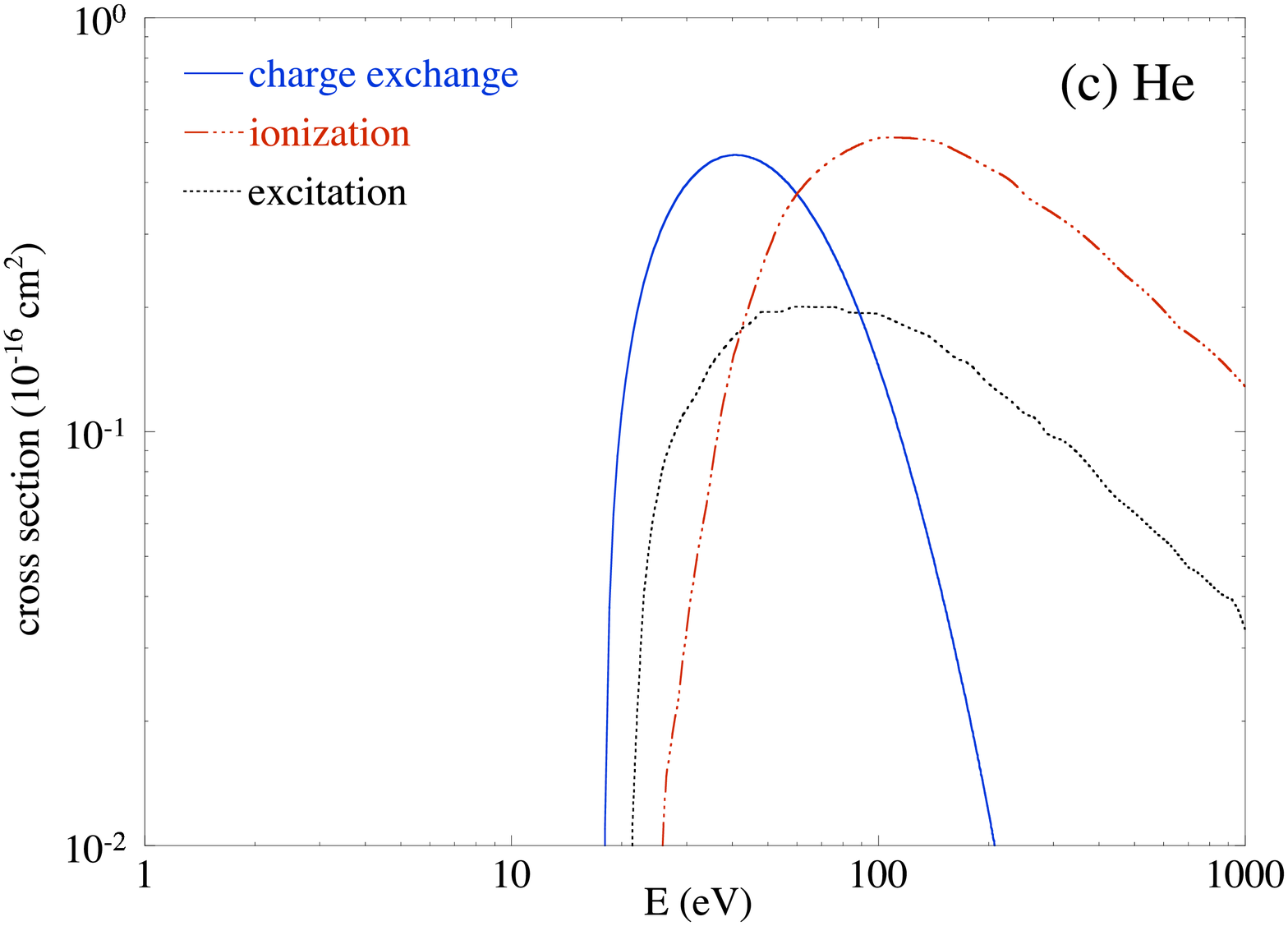}
\caption{Cross sections for positronium formation, ionization and 
excitation in positron collision with (a) atomic hydrogen, (b) molecular 
hydrogen and (c) helium as a function of the kinetic energy. 
\label{fig:cs}}
\end{figure}

In a partially ionized medium, the positron loses its energy by collisions 
with electrons and by interaction with plasma waves, excitation and 
ionization of atoms and molecules. The energy loss in the plasma is 
a continuous process. Its rate, which depends on the energy of the 
positron, the electron temperature and density in the plasma, is 
given by Book and Ali (1975). The other processes require a collision 
with an individual atom or molecule. In 
this case the amount of the energy lost by the positron depends on the 
positron energy and on the type of collision: excitation or ionization. 
The thresholds for all these processes are presented in Table 
\ref{tab:thres}. Figures \ref{fig:cs}a, \ref{fig:cs}b \&\ \ref{fig:cs}c
show the cross sections we used for for excitation, ionization and charge 
exchange in H, H$_{2}$ and He respectively. 

% Table - thresholds 
\begin{table}
     \caption{Energy threshold of reactions induced by positrons. 
\label{tab:thres}}
     \begin{array}[b]{lc}
\noalign{\smallskip}
\hline
\hline
\noalign{\smallskip}
\mbox{Process}  & \mbox{Threshold (eV)} \\
\hline
\noalign{\smallskip}
\mbox{e$^{+}$ + H $\rightarrow$ Ps + H$^{+}$} & \mbox{6.8} \\
\mbox{e$^{+}$ + H $\rightarrow$ e$^{+}$ + e$^{-}$ + H$^{+}$} & \mbox{13.6} \\
\mbox{e$^{+}$ + H $\rightarrow$ e$^{+}$ + H$^{*}$} & \mbox{10.2} \\
\mbox{e$^{+}$ + H $\rightarrow$ e$^{+}$ + H$^{**}$} & \mbox{12.1} \\
\mbox{e$^{+}$ + He $\rightarrow$ Ps + He$^{+}$} & \mbox{17.8} \\
\mbox{e$^{+}$ + He $\rightarrow$ e$^{+}$ + e$^{-}$ + He$^{+}$} & \mbox{24.6} \\
\mbox{e$^{+}$ + He $\rightarrow$ e$^{+}$ + He$^{*}$} & \mbox{21.2} \\
\mbox{e$^{+}$ + H$_{2}$ $\rightarrow$ Ps + H$_{2}^{+}$} & \mbox{8.6} \\
\mbox{e$^{+}$ + H$_{2}$ $\rightarrow$ e$^{+}$ + e$^{-}$ + H$_{2}^{+}$} & \mbox{15.4} \\
\mbox{e$^{+}$ + H$_{2}$ $\rightarrow$ e$^{+}$ + H$_{2}^{*}$} & \mbox{12.0} 
\\
\noalign{\smallskip}
  \hline
\end{array}
\end{table}

%%%%%%%%%%%%%%%%%%%%%%%%%%%%%%%%%%%%%%%%%%%%%%%%%%%%%%%%%%%%%%%%%%%%%%%%%%
%         2.1.1 Hydrogen
%%%%%%%%%%%%%%%%%%%%%%%%%%%%%%%%%%%%%%%%%%%%%%%%%%%%%%%%%%%%%%%%%%%%%%%%%%

\subsubsection{Hydrogen}

The various cross section measurements for positron-impact ionization of 
H we have found in the literature (Jones et al. 1993; Hofmann et al. 1997 
and Kara 1999) are in agreement, except those of Spicher et al. (1990) 
which are thought to have a systematic normalisation error (Charlton and 
Humberston 2001 - hereafter CH2001). On the theoretical side, the results 
of calculations by Janev \&\ Solov'ev (1998) fit the previous measurements 
below 40 eV, whereas those of Ratnavelu (1991) are in good accord with the 
data above 30 eV but slightly overestimate the cross-section in the 
150-300 eV range. 

We have thus reconstructed the cross section dependence on the positron 
kinetic energy by using the calculations of Janev \&\ Solov'ev (1998) 
below 40 eV and averages of the experimental data above this value. The 
energy loss by positrons in such a collision has been approximated by 
BRD79 to be distributed as a Gaussian with a mean of one-fourth of the 
binding energy and a standard deviation of $\approx$ 2 eV. We adopt this 
approximation, which was inspired by the calculations by Omidvar (1965) 
of the ejected electron velocities in electron-hydrogen ionization 
collision.

Concerning the positron-induced excitation of H we used the calculated 
cross sections of Walters (1988) and Kernoghan et al. (1996) for 1s-2s 
and 1s-2p excitations. The cross section for excitation of 
higher energy levels is taken from Fig. 3 of Stein et al. (1998). The 
latter has been deduced by subtracting the elastic, ionization, 
1s-2s and 1s-2p excitation and Ps formation cross sections calculated 
by Kernoghan et al. (1996) from their total calculated cross section. 
The positron energy lost in an excitation to the 2s or 2p levels 
of H is $\approx$ 10.2 eV. We assume the energy loss in excitation of 
the highest levels to be identical to the one for the n=3 level (12.1 eV). 

Among the recent measurements of Ps formation cross sections (Sperber et 
al. 1992, Zhou et al. 1997, Kara et al. 1999), the results of Sperber 
et al. (1992), which were the first to be performed, differ significantly 
(a factor of $~$4/3) from the others and are thought to be incorrect 
(Hoffman et al. 1997). Ps formation cross sections calculated by 
Kernoghan et al. (1996) are in good accord with the experimental 
results of Zhou et al. (1997) and Kara et al. (1999). We thus use 
this calculation's results in modelling the cross section. Since 
the excitation and ionization processes dominate above 100 eV, 
measurements of the Ps formation cross section are subject to 
large uncertainties. For the same reasons, the extrapolation of 
the charge exchange cross section by a power law or by an exponential 
cut does not affect the results of the fraction of Ps formation 
in-flight significantly.

%%%%%%%%%%%%%%%%%%%%%%%%%%%%%%%%%%%%%%%%%%%%%%%%%%%%%%%%%%%%%%%%%%%%%%%%%%
%         2.1.2 Helium
%%%%%%%%%%%%%%%%%%%%%%%%%%%%%%%%%%%%%%%%%%%%%%%%%%%%%%%%%%%%%%%%%%%%%%%%%%

\subsubsection{Helium}

Experiments for the measurement of the single ionization of helium by 
positron
impact (Fromme et al. 1986; Knudsen et al. 1990; Jacobsen et al. 1995; 
Moxom, Ashley \&\ Laricchia 1996) yield general agreement, except for 
the results of Fromme et al. (1986) which slightly overestimate the cross 
section below 50 eV and underestimate it (by $\approx$ 12\%) in the 
100-200 eV range. The calculation for positron kinetic energies below 
150 eV made by Campbell et al. (1998) fits the data quite well 
(see also CH2001 p 241). We use this theoretical result for energies 
less than 150 eV and 
the average of the measurements above this value. The energy distribution 
of the ejected electron is taken from Goruganthu et al. (1985), who 
calculated the single differential cross section of secondary 
electron production for electron-impact ionization in helium. These 
distributions can be approximated by an exponential law with an energy 
scale of $\approx$18.5 eV. We neglected the He double ionization by a
positron since its cross section is $\approx$3  $\times$ 10$^{-3}$ 
lower than that for single ionization (Charlton et al. 1988, CH2001).

The cross section for the excitation of He by positrons is not well 
determined, either by measurements or by calculations. Chapuis et al. 
(1994) deduced it from the difference between the total cross section and 
those of the charge exchange and ionization cross sections. However, the 
excitation cross-section for positron-He scattering evaluated with this 
method overestimates by a factor of $\approx$4 the experimental 
measurements 
(Coleman et al. 1982; Mori \&\ Suoeka 1994) and theoretical calculations 
(Parcell et al. 1987; Hewitt, Noble \&\ Bransden 1992; Ficocelli
Varrachio \&\ Parcell 1992; Campbell et al. 1998; see also Fig. 5.4 
of CH2001). We used the cross section curve of Chapuis et al. (1994) 
scaled to fit the most recent experimental results of Mori \&\ Suoeka 
(1994).

While the measurements of the cross section for positronium formation 
in positron collisions with helium (Fornari, Diana \&\ Coleman 1983; 
Diana et al. 1986; Fromme et al. 1986; Moxom et al. 1993; Overton et 
al. 1993) are in good agreement below $\approx$ 80 eV, large 
discrepancies occur above this value. These discrepancies may be due 
to experimental errors (CH2001), and the most recent results 
of Overton et al. (1993), which are lower than the other measurements 
above 80 eV, are in better accord with theoretical calculations of 
McAlinden \&\ Walters (1992) and Campbell et al. (1998). We have used the 
results of the latter authors to model the cross section for charge 
exchange in helium.

%%%%%%%%%%%%%%%%%%%%%%%%%%%%%%%%%%%%%%%%%%%%%%%%%%%%%%%%%%%%%%%%%%%%%%%%%%
%         2.1.3 Molecular Hydrogen
%%%%%%%%%%%%%%%%%%%%%%%%%%%%%%%%%%%%%%%%%%%%%%%%%%%%%%%%%%%%%%%%%%%%%%%%%%

\subsubsection{Molecular hydrogen}

The four cross section measurements by Fromme et al. (1988), Knudsen 
et al. (1990), Jacobsen et al. (1995) and Moxom et al. (1996) 
of the positron-impact single ionization of molecular hydrogen provide 
very different results for energies less than 200 eV (see also CH2001). 
The latest results of Moxom et al. (1996) seem more appropriate since
they are not far from the results of Fromme et al. (1988) and Knudsen 
et al. (1990). 

However, data of Jacobsen et al. (1995), which underestimate the 
results of previous experiments by $\approx$40\%\ at 80 eV, are the 
closest 
to the theoretical evaluations performed by Chen, Chen \&\ Kuang (1992). 
So by chosing to use the cross section of Moxom et al. (1996), we 
ran the risk of making a systematic error in the calculation 
of the fraction of positronium formation in flight in molecular hydrogen. 
But, as presented in the next section, the results of this assumption 
provide a fraction in agreement with measurements of Brown \&\ Leventhal 
(1986). 

There are few measurements and studies on the energy distribution of the 
ejected electrons in ionisation collision of H$_{2}$ by positrons. 
K\"over \&\ Laricchia (1998), Berakdar (1998) and K\"over, Paludan 
\&\ Laricchia (2002) provide such a distribution, but only for forward 
electrons and specific positron energies. As for helium we approximate 
the distribution of the ejected electrons by an exponential law. 
According to the few existing measurements, the energy scale of this 
distribution is approximately 15 eV. Since the cross section for the 
dissociative ionization of molecular hydrogen by electrons contributes 
only to 6\%\ of the ionization cross section 
(Rapp, Englander-Golden \&\ Briglia 1965), we neglect this 
process in the simulation. 

Measurements of the charge exchange cross section in molecular 
hydrogen are mostly in agreement (Fornari et al. 1983; Diana et al. 
1986; Fromme et al. 1988; Zhou et al. 1997 -- see also Kwan et al. 1998). 
However, the Fromme et al. (1988) results are above the upper 
limit derived by Zhou et al. (1997) for energies less than 10 eV.
Therefore we have used the cross section measured by Fornari 
et al. (1983) and Diana et al. (1986) for the low (E$<$100 eV) and high 
(E$>$100eV) energies, respectively. It should be noted that the 
cross section derived by BRD79 fits the data of Fornari et 
al. (1983) quite well in the 15 -- 50 eV range. 

Very little is known from either theory or experiment concerning the 
excitation of molecules by positrons. 
We thus estimate the excitation cross section of positrons colliding with 
molecular hydrogen by subtracting the charge 
exchange, ionization and elastic cross sections from the total cross 
section, which has been measured by Hoffman et al. (1982). The elastic 
cross section was evaluated by Wallyn et al. (1994).

%%%%%%%% Figure X: Cross sections %%%%%%%%

%%%%%%%%%%%%%%%%%%%%%%%%%%%%%%%%%%%%%%%%%%%%%%%%%%%%%%%%%%%%%%%%%%%%%%%

%%%%%%%%%%%%%%%%%%%%%%%%%%%%%%%%%%%%%%%%%%%%%%%%%%%%%%%%%%%%%%%%%%%%%%%%%%
% 2.2 Positronium formation in flight
%%%%%%%%%%%%%%%%%%%%%%%%%%%%%%%%%%%%%%%%%%%%%%%%%%%%%%%%%%%%%%%%%%%%%%%%%%

\subsection{Positronium formation in flight}

We simulate the interactions of positrons with atoms and molecules 
by Monte-Carlo methods on the basis of the cross sections and energy loss 
mechanisms 
presented previously. For any value of the positron's kinetic energy (that 
is at any stage of the simulation), the interaction process is chosen 
randomly according to the values of the cross sections of all the possible 
processes. The particular interaction then specifies the energy lost at 
that stage. The positron is ``dropped" either when it forms a positronium 
or when its kinetic energy falls below the threshold of charge exchange; 
in the latter case, the positron is assumed to start thermalizing with the 
ambient medium. 

The fraction f$_{1}$ of positrons forming a positronium in flight is then 
obtained simply by counting the number of positrons undergoing a 
charge exchange. When such a process occurs, the kinetic energy value of 
the positronium (i.e. the kinetic energy of the positron minus the 
energy required to form the positronium) is stored. The set of kinetic 
energies 
obtained from the simulation is then used to 
calculate the spectral shape of the annihilation emission by 
accounting for the Doppler broadening due to the motion of 
the positronium. The shapes of the annihilation line differ somewhat from 
a simple Gaussian, but we parametrize them with FWHMs by fitting Gaussian 
functions to the 
spectra (see section 3.2). We also record the time spent by positrons to 
slow down and reach the charge exchange threshold or form a 
positronium. The resulting mean duration deduced from a large number 
of positrons allows us to calculate the slowing down time of positrons 
before thermalization.

\begin{figure}[tb]
%\centering
\includegraphics[width=8.cm,height=8.cm]{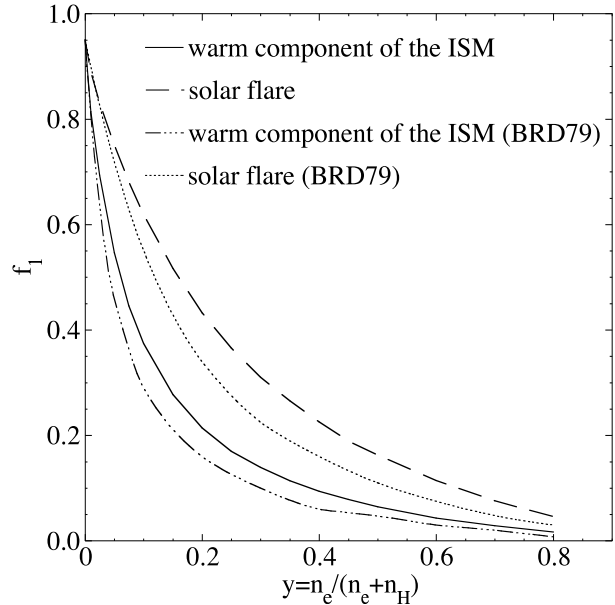}
\caption{Fraction of positrons forming positronium in flight 
(f$_{1}$) by charge exchange with atomic hydrogen as a function of the 
ionization fraction (y) in a warm component of the interstellar medium 
(electron density: n$_{e}$ = 0.1 cm$^{-3}$, electron temperature: 
T$_{e}$ = 8000 K) and 
in solar flare (n$_{e}$ = 5 $\times$ 10$^{13}$ cm$^{-3}$, T$_{e}$ = 
1.16 $\times$ 10$^{4}$ K). The results presented in Fig. 2 of BRD79 
are shown for comparison.
\label{fig:brd79}}
\end{figure}

We tested our Monte Carlo simulation by reproducing the results 
obtained by BRD79 (see Figure \ref{fig:brd79}). They 
calculated the fraction of positrons undergoing charge exchange with 
atomic hydrogen before thermalization as a function of the ionization 
fraction. Similarly our calculation has been performed for two cases 
of media characterized by their electron density and temperature. The 
first is representative of a warm component of the ISM (T~$\sim 8000$ K) 
and the second of solar flares (T~$\sim 10^6$ K).

%%% Figure : f1 vs ionization fraction %%%
%%%%%%%%%%%%%%%%%%%%%%%%%%%%%%%%%%%%%%%%%%%%%%%%%%%%%%%%%%%%%%%%%%%%%%%

Values of f$_{1}$ for a totally neutral medium are summarized in Table 
\ref{tab:f1} where they are compared with results of previous 
measurements (Brown \&\ Leventhal 1984, 1986) and calculations 
(BRD79; Wallyn et al. 1994; Chapuis et al. 1994).

%%%%%%%% Table - comparaison des differents f1 obtenus %%%%%%%%%%%%%%%%
\begin{table}
\caption{Fraction (in \%) of positrons forming positronium in flight, 
in a completely neutral medium.\label{tab:f1}}
\begin{array}[b]{lccc}
\noalign{\smallskip}
\hline
\hline
\noalign{\smallskip}
\mbox{References}  & \mbox{H} & \mbox{H$_{2}$} & \mbox{He} \\
\hline
\noalign{\smallskip}
\mbox{BRD79} & 95 & 93 & - \\
\mbox{Brown \& Leventhal} \qquad & - & \quad 89.7 \pm 0.3 \quad & 80.7 \pm 0.5 \\
\mbox{Wallyn et al. 1994} & 98 & 90 & - \\
\mbox{Chapuis et al. 1994} \qquad & - & - & 78 \\
\mbox{This paper} & 95.5 & 89.6 & 81.7 \\
\noalign{\smallskip}
  \hline
\end{array}
\end{table}
%%%%%%%%%%%%%%%%%%%%%%%%%%%%%%%%%%%%%%%%%%%%%%%%%%%%%%%%%%%%%%%%%%%%%%%

%%%%%%%%%%%%%%%%%%%%%%%%%%%%%%%%%%%%%%%%%%%%%%%%%%%%%%%%%%%%%%%%%%%%%%%%%%
% 3. Processes after Thermalization
%%%%%%%%%%%%%%%%%%%%%%%%%%%%%%%%%%%%%%%%%%%%%%%%%%%%%%%%%%%%%%%%%%%%%%%%%%

\section{Processes after thermalization}

After they have lost the bulk of their energies during the ``in-flight" 
phase, the positrons thermalize with the ambient medium and undergo a 
series of processes leading to their annihilation, either directly 
or via positronium formation.

%%%%%%%%%%%%%%%%%%%%%%%%%%%%%%%%%%%%%%%%%%%%%%%%%%%%%%%%%%%%%%%%%%%%%%%%%%
% 3.1 Reaction rates for positron processes
%%%%%%%%%%%%%%%%%%%%%%%%%%%%%%%%%%%%%%%%%%%%%%%%%%%%%%%%%%%%%%%%%%%%%%%%%%

\subsection{Reaction rates for positron processes}

In thermal conditions, the reaction rate for non-relativistic binary 
collisions is easy to calculate; we use the usual $<\sigma v>$ formula, 
where the averaging is done over the interacting particles' (Maxwellian) 
energy distributions:

\begin{equation}
 R = <\sigma v> = \int_{E_T}^{\infty}{2 \over \sqrt{\pi}} 
{{\sqrt{E}\over 
{(kT)}^{3/2}} e^{-E/kT} \sigma(E) v dE} \; . 
\label{eq:therm} 
\end{equation}

%%%%%%%% Figure X: Reaction Rates as a function of Temperature %%%%%%%%
\begin{figure}[tb]
%\centering
\includegraphics[width=1.0\linewidth ,height=9.0cm,width=8.5cm]{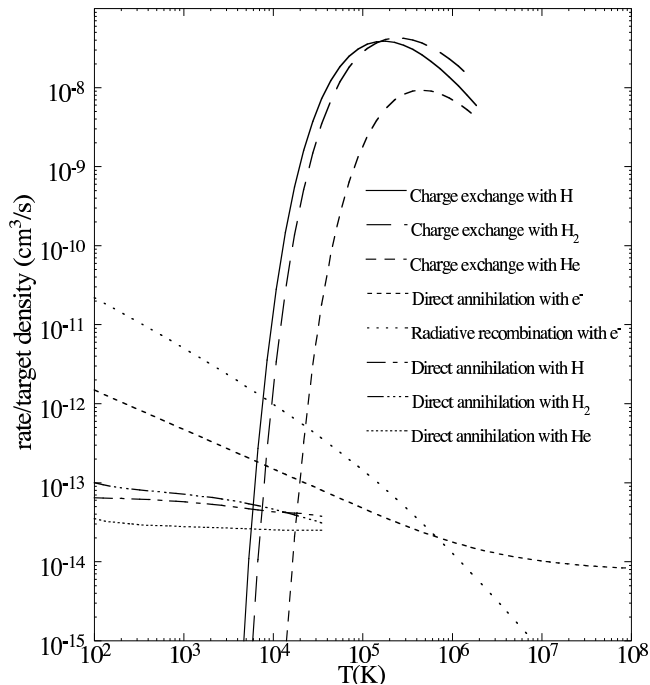}
\caption{Positron reaction rates as a function of temperature. 
\label{fig:Rates}}
\end{figure}

%%%%%%%%%%%%%%%%%%%%%%%%%%%%%%%%%%%%%%%%%%%%%%%%%%%%%%%%%%%%%%%%%%%%%%%

%%%%%%%%%%%%%%%%%%%%%%%%%%%%%%%%%%%%%%%%%%%%%%%%%%%%%%%%%%%%%%%%%%%%%%%%%%
%         3.1.1 charge exchange with H, H$_{2}$, He
%%%%%%%%%%%%%%%%%%%%%%%%%%%%%%%%%%%%%%%%%%%%%%%%%%%%%%%%%%%%%%%%%%%%%%%%%%

\subsubsection{Charge exchange with H, H$_{2}$, He}

The cross sections for these reactions were discussed in the previous 
section. The calculated rate is shown in Figure \ref{fig:Rates} along with 
the rates of the other processes. These rates are significant only above 
a few times 10$^{4}$ K, when the high-energy tail of the Maxwellian 
distribution is above the threshold of Ps formation. Also, although 
few atoms remain neutral at such a high temperatures, we present the 
charge exchange rate up to 2 $\times$ 10$^{6}$ K.

%%%%%%%%%%%%%%%%%%%%%%%%%%%%%%%%%%%%%%%%%%%%%%%%%%%%%%%%%%%%%%%%%%%%%%%%%%
%         3.1.2 radiative combination (with free electrons)
%%%%%%%%%%%%%%%%%%%%%%%%%%%%%%%%%%%%%%%%%%%%%%%%%%%%%%%%%%%%%%%%%%%%%%%%%%

\subsubsection{Radiative combination (with free electrons)}

The cross section for this process (e$^+$ + e$^-$ $\rightarrow$ Ps $ + 
\gamma$) is so small ($~ 10^{-20}$ cm$^2$ at 1 eV and decreasing fast at 
higher energies) that it cannot be measured in the laboratory. We must 
thus rely on theoretical calculations. Various attempts and approaches 
have been made over the past several decades, going back to 1930, the most 
recent and accurate of which is that of Gould (1989), who obtained the 
reaction rate for both the ground state and the first excited state of Ps 
directly from first principles. The earlier calculation of Crannell et al. 
(1976) used the similarity between this process and the radiative 
recombination of hydrogen (e$^-$ + p $\rightarrow$ H $ + \gamma$) for 
which the cross section was already known.

%%%%%%%%%%%%%%%%%%%%%%%%%%%%%%%%%%%%%%%%%%%%%%%%%%%%%%%%%%%%%%%%%%%%%%%%%%
%         3.1.3 direct annihilation with free electrons
%%%%%%%%%%%%%%%%%%%%%%%%%%%%%%%%%%%%%%%%%%%%%%%%%%%%%%%%%%%%%%%%%%%%%%%%%%

\subsubsection{Direct annihilation with free electrons}

This process (e$^+$ + e$^-$ $\rightarrow \gamma + \gamma$) has a very 
small cross section (about an order of magnitude less than that of 
radiative 
combination at temperature less than $~ 10^5$ K -- Heitler 1954; 
Crannell et al. 1976; BRD79). It is therefore only relevant in the hot 
phase 
of the ISM and can be ignored in all other conditions, as can be see in 
Figure \ref{fig:Rates}.

%%%%%%%%%%%%%%%%%%%%%%%%%%%%%%%%%%%%%%%%%%%%%%%%%%%%%%%%%%%%%%%%%%%%%%%%%%
%         3.1.4 direct annihilation with bound electrons
%%%%%%%%%%%%%%%%%%%%%%%%%%%%%%%%%%%%%%%%%%%%%%%%%%%%%%%%%%%%%%%%%%%%%%%%%%

\subsubsection{Direct annihilation with bound electrons}

This process (e$^+$ + H $\rightarrow p + \gamma + \gamma$, or e$^+$ on 
H$_2$ or on He) has the lowest 
cross section of all the positron processes, more 
than an order of magnitude less than the second slowest process, direct 
annihilation with free electrons, and more than two orders of magnitude 
weaker than radiative combination. Direct 
annihilation with bound electrons applies only at the lowest temperatures 
(less than a few thousand degrees) where charge exchange does not proceed 
because the positrons then do not have the threshold energy needed
and there are no free electrons for the other two processes -- radiative 
combination and direct annihilation with free electrons -- to take place. 
Due to the weakness of its cross section, this process cannot be 
investigated in the laboratory, and again one must rely on theoretical 
calculations. The work of Bhatia, Drachman and Temkin (1977) has not been 
corrected or updated by any later works, so we still use the rate they 
obtained for direct 
annihilation with H. For He and H$_{2}$ the method is similar, except that 
we calculated the cross section using estimates of the dependence of 
Z$_{eff}$, the effective charge\footnote{The effective charge 
Z$_{eff}$ is defined such that the annihilation rate per 
target density is $R = \sigma v = Z_{eff} c 
\pi r_e^2$, with $r_{e}$ the classical electron radius.},
on the kinetic energy from Campeanu \&\ Humberston (1977) and 
Armour, Baker \&\ Plummer (1990), respectively. 

The measured Z$_{eff}$ in H$_{2}$ at room 
temperature is $\approx$45\%\ larger than the theoretical evaluation 
(CH2001). This may turn out to be a source of systematic error in the 
annihilation rate we will be using.

%%%%%%%%%%%%%%%%%%%%%%%%%%%%%%%%%%%%%%%%%%%%%%%%%%%%%%%%%%%%%%%%%%%%%%%%%%
% 3.2 Line widths
%%%%%%%%%%%%%%%%%%%%%%%%%%%%%%%%%%%%%%%%%%%%%%%%%%%%%%%%%%%%%%%%%%%%%%%%%%

\subsection{Line widths}

The widths of the annihilation line resulting from charge exchange in 
flight with H, H$_2$, and He have been calculated as a result of the 
Monte Carlo simulation we presented in the previous section. 
We have found 5.8 keV, 6.4 keV and 7.4 keV for H, H$_2$ and He 
respectively in a neutral medium. These widths are in quite good 
agreement with previous measurements and calculations (BRD79, Brown 
\& Leventhal 1984, Brown \& Leventhal 1986). The width of the 
511 keV line does not vary with the ionisation fraction of the ambient 
medium although its shape is slightly ``flattened" (the top of the 
spectral distribution is less peaked). This flattening is 
due to the loss of low energy positrons which are quickly slowed down 
below the Ps threshold by way of Coulomb collisions with electrons. 

The widths of the annihilation line resulting from the direct annihilation 
with H, H$_2$, and He have all been measured experimentally: Brown \& 
Leventhal (1986) obtained 1.56 keV for H; and Iwata, Greaves, and Surko 
(1997) obtained 1.71 keV and 2.50 keV for H$_2$ and He, respectively. 
These widths depend very weakly on the temperature in our range of 
interest.

The widths of both the radiative combination and direct annihilation with 
free electrons were calculated by Crannell et al. (1976) using a simple 
argument of thermal broadening due to the motion of the center of mass of 
the e$^+$-e$^-$ pair; this calculation gives the same following simple 
expression for both widths: $\Gamma_{rc, dae} = 1.1$ T$_4^{1/2}$ keV, 
where T$_4$ is the temperature in units of $10^4$ K. 

%%%%%%%%%%%%%%%%%%%%%%%%%%%%%%%%%%%%%%%%%%%%%%%%%%%%%%%%%%%%%%%%%%%%%%%%%%
% 3.3 Grains and positron processes
%%%%%%%%%%%%%%%%%%%%%%%%%%%%%%%%%%%%%%%%%%%%%%%%%%%%%%%%%%%%%%%%%%%%%%%%%%

\subsection{Grains and positron processes}

The importance of positron capture and annihilation on interstellar dust 
grains was first pointed out by Zurek (1985), who estimated the rate for 
the process and concluded that dust can play a crucial role in the warm 
ionized phase of the ISM, leading to drastic modifications of the Ps 
formation fraction depending on the value of the cross section of the 
positron capture on the dust grains. Guessoum, Ramaty and Lingenfelter 
(1991) added spectral considerations to the positron-dust process but kept 
a simplified picture of the dust and showed that unless the abundance of 
dust in the annihilation regions is increased many times compared to 
``normal" quantities, grains change the spectrum of the line mostly in the 
hot phase. Astrophysical knowledge of grains has since increased 
considerably (see the reviews by Draine 2003, 2004, Li 2004, and Krugel 
2003) and although a huge body of work exists on positron interactions 
with solids and solid surfaces (Nieminen \& Oliva 1980; Huttunen et al. 
1990; Dannefaer et al. 1996; Dupasquier \& Mills 1995; Puska \& Nieminen 
1994; Weiss et al. 1994; Suzuki et al. 2003; Mills et al. 1989; Mokrushin 
et al. 2003; Pi et al. 2003; Weber \& Lynn 2000), experiments 
dealing with positrons and dust like materials remain 
scarce. We here summarize our best understanding of the 
relevant information we have collected on dust grains in our context.

The total amount of dust in the Galaxy is estimated at $3 \times 10^7 
M_\odot$, which represents 0.6 \% of the total mass of gas (Krugel 2003), 
although Ferriere (2003) indicates that there is a large uncertainty on 
this fraction, 
which could range from 0.5 to 1 \%. The abundance 
of dust grains is found to correlate approximately with hydrogen, 
particularly in the cold and neutral phases; there is also some evidence 
(Herter et al. 1989; Krugel 2003) for partial dust destruction in the HII 
regions.

Several species of dust (composition, size) exist in the ISM. Dust is 
often 
divided into 3 categories as follows:

\begin{itemize}

\item Large grains, of radii ranging between 20 nm and 0.3 $\mu$m, made 
of either silicate particles (Mg$_2$SiO$_4$ or Fe$_2$SiO$_4$) or amorphous 
carbon;

\item Very small grains, of radii ranging between 1 nm and 10 nm, made 
mostly 
of graphite, and representing about 10 \% of the amorphous carbons;

\item PAH (polycyclic aromatic hydrocarbons), which are big molecules and 
make up 3 to 6 \% of the large grains (in abundance).

We have ignored the PAHs in the present treatment due to their negligible 
contribution; in fact, we have found that the ``very small grains" can be 
neglected as well, as their abundance and small sizes (small geometric 
cross sections) give extremely low positron capture rates.

\end{itemize}

Grain sizes follow a power-law distribution: dn$_{gr} \propto a^{-3/2} 
da$, where $a$ is the radius of the grain, which ranges between $3 \times 
10^{-6}$ cm and $2.4 \times 10^{-5}$ cm for the large carbonaceous grains 
and between $1.5 \times 10^{-6}$ cm and $1.2 \times 10^{-5}$ cm for the 
large silicate grains (Krugel 2003). Other distributions and grain size 
ranges that distinguish between grain types have also been proposed 
(Weingartner \& Draine 2001; Li \& Draine 2001), but we will adopt the 
simple unified expression of Krugel (2003). 
The distributions and grain size ranges given by Weingartner \& Draine 
(2001) or Li \& Draine (2001) lead to reaction rates that are several 
times higher than those we obtain with the ``conservative" values of 
Krugel (2003). This factor (the sizes and distribution law of the grains) 
is the fundamental uncertainty in the determination of the reaction rate 
of positron interaction with the dust. Secondary factors include the 
amount of dust in the ISM (as stated previously) and the electric charge 
of the grains in the cold phase. We have combined these uncertainties 
into one parameter, denoted by $x_{gr}$ (similar to that of Guessoum, 
Ramaty and Lingenfelter 1991), where $x_{gr} = 1$ represents our 
``standard" rate, obtained on the basis of the Krugel (2003) 
parametrization of dust; $x_{gr} = 0$ represents a total absence of dust 
in a given region; $x_{gr} > 1$ represents an overabundance of dust and/or 
grain characteristics (sizes and composition) that are especially 
favorable for positron capture. We emphasize however, that these factors do 
not affect our conclusions, namely that the reaction rates for 
positron capture on dust are, even in the most extreme assumptions, 
negligible compared to other processes' in all but the warm-ionized and 
hot ISM phase: in the latter environment the dust dominates all other 
processes under all assumptions, and in the other case the dust becomes 
comparable or dominant only when the grain capture rate 
(or equivalently $x_{gr}$) is increased by a large factor.

Positron collisions with dust grains can be treated along the same lines 
as Zurek (1985) and Guessoum, Ramaty and Lingenfelter (1991) did, namely 
that the cross section is essentially a geometric one ($\pi$ a$^2$) with 
the electric charge effect (attraction or repulsion with the positron 
depending on the local conditions) and the positron escape (backscattering 
and re-emission from the dust grain) taken into account:

\begin{equation}
\sigma_{gr} = (1 - R_+) \pi a^2 f_{elec} \; , 
\label{eq:siggr}
\end{equation}

\noindent where a is the radius of the grain and R$_+$ is the probability 
that the positron is either backscattered by the grain or re-emitted after 
being captured. f$_{elec}$ is given by (1 - Ze$^2$/akT) or 
exp(-Ze$^2$/akT) 
depending on whether the grain is negatively or positively charged, 
respectively; Z is the grain charge, and T is the temperature of the 
medium. Grains are expected to be negatively charged in the hot and 
ionized phases of the ISM and possibly positively charged in the cold and 
neutral phases (being slightly ionized themselves by the UV radiation of 
nearby stars if applicable). It can be shown (Krugel 2003) that the 
electric potential of a grain is 2.5 kT/e, so that the quantity Ze$^2$/akT 
= 2.5 is independent of the temperature of the medium or the size of the 
grain.

The positrons that are captured will diffuse and thermalize quickly with 
the grain molecules; infra-red emission of the ISM dust shows that the 
temperature of the grains varies between $~ 15$ K and 100 K depending on 
the conditions of the ISM medium (15K in the coldest phases, and between 
50 K and 100 K in the ionized and hot conditions). Once thermalized, the 
positrons in the grains will form a positronium atom, which will either 
readily be in the para-state or be formed in the ortho-state but undergo a 
``pick-off" annihilation reaction as a para-state (giving 2 photons) when 
the positron (bound in the Ps) ``picks off" an electron from the 
surrounding valence electrons. Most of the positrons 
that diffuse inside the grain will reach the surface, the diffusion length 
being greater than the diameter of most of the grains, but upon reaching 
the surface the positron will be captured in a ``surface state" and form 
Ps there or be ejected, if the ``positron work function" of the 
material is negative, which is the case for our dust grains 
(Rosenberg, Howell \&\ Fluss 1987). A 
fraction of the Ps atoms formed at or near the surface will also be 
ejected if the ``Positronium work function" for the material is negative, 
which is also the case for our dust (Hodges \&\ Stott 1973, Nieminem 
\&\ Oliva 1980, Eldrup et al. 1985, Rice-Evans \&\ Rao 1988, Puska \&\ 
Nieminem 1994). 

Aside from the backscattering probability measured and simulated for Si 
and a few other elements (Makinen et al. 1992), we have found no specific 
measurements of e$^+$ and Ps probabilities for the materials that 
constitute our dust grains in the conditions that we deal with (slow 
positrons, etc.). There are some experiments (and a few calculations) that 
deal with metals (Huttunen et al. 1990; Nieminen \& Oliva 1990) and 
polymers (Mukherjee, Chakravorty and Nambissan 1998) or conduct studies 
(probing) of defects in semi-conductors by positrons (Britton, Hempel and 
Triftshauser 2001; van Veen et al. 2004), but none that consider materials 
of the dust type.

We have adopted the value of 10 \% of positron backscattering probability 
obtained by Makinen et al. (1992) when they extrapolate to the lowest 
positron energies (less than 1 keV), a value that seems to apply for all 
the materials they investigated (including Si). 

For the rest of the parameters, we can only use measurements 
for materials that have similar chemical behavior (positron and 
Positronium negative work functions), and the closest we could find were 
copper and aluminum. For instance, the probability of re-emission of 
positrons from copper varies linearly with T from 0 to 15 \% between 0 and 
150 K, the temperature of the metal, which we will take as the temperature 
of the dust, in our case. (Nieminen \& Oliva's 
theoretical treatment predicted that this probability would tend to zero 
at very low temperatures.) The probability of ejection of Ps from the 
surface of the grain is found to be very similar for Cu and Al, taking the 
following approximate values: 0 at 0 K, 5 \% at 25 K, 9 \% at 50 K, 14 \% 
at 100 K, 17 \% at 150 K and 20 \% at 200 K.

If these measurements are used for our dust grains (whose temperature 
ranges between 15 and 100 K), it would imply that the bulk (65 to 80 \%) 
of the positrons colliding with grains will annihilate inside (giving two 
photons), a small fraction (5 to 15 \%) will form Ps, be ejected 
and annihilate outside (giving two/three photons in the usual 25/75 \% 
proportions), while 15 to 20 \% of the positrons will simply fly back out 
after hitting a grain.

\begin{figure*}
    \begin{center}
\includegraphics[width=12cm,height=8cm]{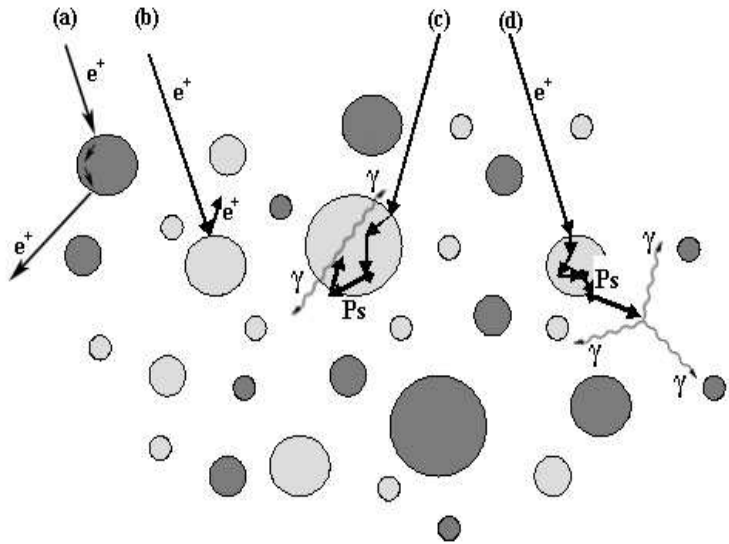}
\caption{Collisions and subsequent annihilation of positrons with grains 
(of various sizes and types) - (a) positron re-emission, (b) positron 
backscattering, (c) Positronium in grain, (d) Positronium ejection.
\label{fig:grain}}        
    \end{center}
\end{figure*}

Figure \ref{fig:grain} gives a schematic description of the main processes 
that a positron can undergo during its interaction(s) with dust grains.

%%%%%%%% Figure XY: Positrons-Grains %%%%%%%%
%\begin{figure}[tb]
%\centering
%%%%%%%%%%%%%%%%%%%%%%%%%%%%%%%%%%%%%%%%%%%%%%%%%%%%%%%%%%%%%%%%%%%%%%%

The rate $r_{gr}(a)$ of positron capture by grains of size a is obtained from the 
effective cross section of positron capture by grains and the thermal 
speed of positrons in the medium:

\begin{equation}
r_{gr}(a) = n_{gr} \, \sigma_{gr} \, v_{e^+} \, , 
\label{eq:rggr1}
\end{equation}

\noindent where $n_{gr}$ is the number density of dust grains in the 
ISM, $\sigma_{gr}$ is the cross section of positron-capture 
by grains (Equation \ref{eq:siggr}, and $v_{e^+}$ is the average thermal 
speed of positrons in the given ISM phase: $v_{e^+} = \sqrt{{3kT}\over 
m_e}$. 
$n_{gr}$ is obtained from the overall mass of dust in the galaxy (0.6 \% 
of 
the total gass mass) as well as the the grain size distribution 
$ dN_{gr} = C_0 a^{-3/2} da$, which yields the mean value of the 
grain's mass ($<m_{gr}>_{a} = \int_{a_-}^{a_+} \rho_{gr} 4\pi /3 a^3 
dN_{gr}$). 
The constant C$_0$ is inferred from the normalization condition 
$ \int dN_{gr} = 1$. For the carbonaceous dust, the density is 
$ \rho_{gr, C} \approx $ 2.24 g cm$^{-3}$; for the silicate dust, 
$ \rho_{gr, Si/Mg/Fe} \approx $ 3.5 g cm$^{-3}$. A simple calculation, 
assuming the total dust mass to be equally divided into the two main 
types, then gives: $n_{gr} \approx 3.5 \times 10^{-13} n_H$ for the 
carbonaceous dust and $ 1.8 \times 10^{-12} n_H$ for the silicate dust.

Replacing $\sigma_{gr}$ by its expression (Equation \ref{eq:siggr}) 
in Equation \ref{eq:rggr1} allows one to show explicitly the factors 
that influence 
the value of the positron capture rate: the reflection factor R$_+$ of positrons 
backscattered or ejected from the dust grain as explained above, the 
electrostatic enhancement of inhibition of positron-grain collision 
f$_{elec}$. We add an additional factor f$_{dest}$ to take into account 
the potential destruction of dust grains, most likely 
in the hot environments. Averaging over the grain size distribution, 
the rate then becomes:

\begin{equation}
r_{gr} = n_{gr} \, (1 - R_+) f_{elec} f_{dest} <\sigma_{gr,geom}>_a v_{e^+} \, ,
\label{eq:rggr2}
\end{equation}

where $\sigma_{gr,geom} = \pi a^2$ is the geometric cross-section 
and $<\sigma_{gr,geom}>_a$ the mean value of the geometric cross section. 
As stated previously, dust is likely to be 
partially destroyed only in the hot phase of the ISM, so we take 
$f_{dest}$ = 0.5 in that case, and $f_{dest}$ = 1 in all other 
conditions. In the hot and warm-ionized phases, grains are widely 
considered to be negatively charged, with a potential such that 
$Ze^2/akT \approx -2.5$, so that $f_{elec} = 1 - Ze^2/akT \approx 3.5$; 
in the cold and neutral phases, however, the grains are assumed 
to be positively charged, but their electrostaic potential 
depends on local conditions (proximity of an 
ionizing UV radiation from a neighboring hot star, etc.), so that the 
potential can vary between 0 and 2.5 V; in these cases, and considering the 
de facto impossibility of pinpointing such a value we take $f_{elec} = 
exp(-Ze^2/akT) \approx 0.5$, keeping in mind that such a factor is in 
effect irrelevant in the cold phase since the dust has no effect on the 
annihilation there. 

Taking all these factors into account then yields final values for 
the positron capture rates by the carbonaceous and the silicate dust 
for the various ISM phase conditions, as reported in Table \ref{tab:RR}. 

%%%%%%%%%%%%%%%%%%%%%%%%%%%%%%%%%%%%%%%%%%%%%%%%%%%%%%%%%%%%%%%%%%%%%
% Table - Reaction rates for various processes in the different phases
%
\begin{table*}
\caption{Reaction rates (in cm$^3$/s) for the various processes 
after thermalization in the 
different ISM phases.\label{tab:RR}}
\begin{array}[b]{lccccc}
\noalign{\smallskip}
\hline
\hline
\noalign{\smallskip}
\mbox{Process / Medium} & 
\mbox{Molecular }& 
\mbox{Cold }& 
\mbox{Warm Neutral }& 
\mbox{Warm Ionized }& 
\mbox{Hot } \\
& \mbox{(T $\approx$ 10 K)} & \mbox{(T $\approx$ 80 K)} & \mbox{(T $\sim $ 
8000 K)} & \mbox{(T $\sim $ 8000 K)} & \mbox{(T $\sim $ 10$^6$ K)} \\
\hline
\noalign{\smallskip}
\mbox{Charge Exchange with H} & -- & -- & \mbox{1.8 $\times$ 10$^{-12}$} & -- & -- \\
\mbox{Charge Exchange with H$_2$} & -- & -- & -- & -- & -- \\
\mbox{Charge Exchange with He} & -- & -- & \mbox{9.0 $\times$ 10$^{-21}$} & \mbox{9.0 $\times$ 10$^{-21}$} & -- \\
\mbox{Direct Annihilation with H} & -- & \mbox{6.5 $\times$ 10$^{-14}$} & \mbox{4.4 $\times$ 10$^{-14}$} & -- & -- \\
\mbox{Direct Annihilation with H$_2$} & \mbox{4.3 $\times$ 10$^{-13}$} & -- & -- & -- & -- \\
\mbox{Direct Annihilation with He} & \mbox{1.5 $\times$ 10$^{-13}$} & \mbox{3.7 $\times$ 10$^{-14}$} & \mbox{2.6 $\times$ 10$^{-14}$} & \mbox{2.6 $\times$ 10$^{-14}$} & -- \\
\mbox{Radiative Combination} & -- & -- & -- & \mbox{1.2 $\times$ 10$^{-12}$} & \mbox{1.3 $\times$ 10$^{-14}$} \\
\mbox{Direct Annihilation with electrons}  \quad & \quad -- \quad & \quad -- \quad & \quad -- \quad & \quad \mbox{1.7 $\times$ 10$^{-13}$} \quad & \quad \mbox{1.8 $\times$ 10$^{-14}$} \quad \\
\mbox{Capture by grains} & \mbox{2.4 $\times$ 10$^{-16}$} & \mbox{6.8 $\times$ 10$^{-16}$} & \mbox{6.5 $\times$ 10$^{-15}$} & \mbox{4.6 $\times$ 10$^{-14}$} & \mbox{2.4 $\times$ 10$^{-13}$} \\
\noalign{\smallskip}
  \hline
\end{array}
\end{table*}
%%%%%%%%%%%%%%%%%%%%%%%%%%%%%%%%%%%%%%%%%%%%%%%%%%%%%%%%%%%%%%%%%%%%%

The rates obtained and reported here are in most cases essentially the 
same as those of Guessoum, Ramaty and Lingenfelter (1991), but in some 
cases and due to the use of new data, the rates are substantially 
different. The differences are found in the rates of charge exchange with 
H in the warm neutral phase, where our new value is greater than the 
previous one by about 33 \%; direct annihilation with H in the warm 
neutral phase, where our rate is greater than the old one by about 16 \%; 
capture by grains, where our rates are about a factor of 10 lower, a factor 
of 3 lower and a factor of 1.5 greater than the old values in the cold, 
warm neutral and hot phases respectively.

The width of the annihilation line resulting from the decay of the Ps 
produced inside the grain is an important quantity. There are actually two 
widths, corresponding to the annihilation of the Ps atoms that escape and 
those that remain and die in the grains. Nieminen \& Oliva (1980) state 
that the kinetic energy of the Ps ejected by the grain is equal to the 
absolute value of the ``Ps work function" $\phi_{Ps}$ of the material. 
This implies a width $ \Gamma_{gr, out} = 2 \sqrt{m_ec^2\phi_{Ps}}$ of 
about 1.4 keV, taking $\phi_{Ps} \approx 1$ eV. (Note that the value used 
for $ \Gamma_{gr, out}$ since Guessoum, Ramaty and Lingenfelter 1991 was 
2.5 keV. Moreover, the fraction of Ps escaping from the grain is now taken 
to be 5 - 15 \%, compared to the previously held 1/3 value. This will have 
some significant impact on the shape of the annihilation line, at least in 
the hot and ionized phases.) The width of the line produced by para-Ps 
decaying inside the grain cannot be estimated via other physical 
quantities and needs to be measured in the laboratory. Again, there have 
been no measurements for the kinds of materials that make up our grains, 
but typical ACAR (angular correlation of the annihilation radiation) 
measurements of the two emitted photons give values $\theta_{ACAR} ~ 5 - 8 
$ mrad (Biasini et al. 2000; Sasaki et al. 2003), which would translate 
into $ \Gamma_{gr, in} = 0.5 \theta_{ACAR}m_ec^2 \approx 1.3 - 2.0$ keV. 
We note, however, that the measurements of $ \Gamma_{gr, in}$ by Iwata, 
Greaves and Surko (1997) for many atoms and molecules (including complex 
alcohols) shows the width to be always between 2.0 keV and 3.0 keV. We 
have thus adopted the higher value of the ACAR measurements, which 
corresponds to the lowest value of the annihilation on molecules: $ 
\Gamma_{gr, in}\approx 2.0$ keV.

%%%%%%%%%%%%%%%%%%%%%%%%%%%%%%%%%%%%%%%%%%%%%%%%%%%%%%%%%%%%%%%%%%%%%%%%%%
% 4. Annihilation Spectra
%%%%%%%%%%%%%%%%%%%%%%%%%%%%%%%%%%%%%%%%%%%%%%%%%%%%%%%%%%%%%%%%%%%%%%%%%%

\section{\label{s4}Annihilation spectra}

Having reviewed and recalculated the various processes' reaction rates 
and line widths, we can construct the standard spectra of the annihilation 
radiation for each phase of the ISM as well as an overall ``global" 
or ``diffuse" radiation spectrum, one that includes contributions from 
the various regions; the latter ``global" spectrum requires knowledge 
about the relative distribution of the ISM phases and their volume 
contributions to the emitted radiation. 

The emission processes are spectroscopically modeled as either a pure 
two-photon direct annihilation represented by a Gaussian with a given 
FWHM or a combined two-photon-line/three-photon-continuum when the 
annihilation proceeds via formation of a positronium atom; in this 
latter case, the two-photon line is represented by a Gaussian with 
its characteristic FWHM, while the continuum part is represented by 
the usual Ore \& Powell (1949) function $P_t(E)$ that describes 
the probability of the emitted photon being emitted with an energy 
E between 0 and 511 keV (Eq.~10 of Guessoum, Ramaty and Lingenfelter, 
1991). 

Table \ref{tab:FWHM} summarizes the line widths for various processes. 
For several processes the Gaussian function with a given 
FWHM represents a slight simplification of the real spectrum, but the 
annihilation spectra obtained in the end, and their eventual comparison 
with the data, are insensitive to this level of precision.

%%%%%%%%%%%%%%%%%%%%%%%%%%%%%%%%%%%%%%%%%%%%%%%%%%%%%%%%%%%%%%%%%%%%%
% Table - Line Widths for various processes 
\begin{table*}

\caption{Line widths (in keV) for various positron annihilation 
processes.\label{tab:FWHM}}
\begin{array}[b]{lcccc}
\noalign{\smallskip}
\hline
\hline
\noalign{\smallskip}
\mbox{Process / Medium} & 
\mbox{Cold \& Molecular}& 
\mbox{Warm Neutral }& 
\mbox{Warm Ionized }& 
\mbox{Hot } \\
& \quad \mbox{(T $<$ 100 K)} \quad & \quad \mbox{(T $\sim $ 8000 K)} \quad 
&  \quad \mbox{(T $\sim $ 8000 K)}  \quad & \quad \mbox{(T $\sim $ 10$^6$ 
K)} \quad \\
\hline
\noalign{\smallskip}
\mbox{Charge Exchange with H in-flight} & 5.8 & 5.8 &  --& -- \\
\mbox{Charge Exchange with H$_2$ in-flight} & 6.4 & -- & --& -- \\
\mbox{Charge Exchange with He in-flight} & 7.4 & 7.4 & 8.7 & -- \\
\mbox{Charge Exchange with H after thermalization} \quad & -- & 1.16 & --& -- \\
\mbox{Charge Exchange with H$_2$ after thermalization} \quad & --& -- & --& -- \\
\mbox{Charge Exchange with He after thermalization} \quad & --& 1.22 & 1.22 & -- \\
\mbox{Direct Annihilation with H}& 1.56 & 1.56 & --& -- \\
\mbox{Direct Annihilation with H$_2$}& 1.71 & -- & -- & -- \\
\mbox{Direct Annihilation with He}& 2.50 & 2.50 & 2.50 & -- \\
\mbox{Radiative Combination}& -- & -- & 0.98 & 11 \\
\mbox{Direct Annihilation with electrons}& --& --& 0.98 & 11 \\
\mbox{Positronium from grains}& 1.4 & 1.4 & 1.4 & 1.4 \\
\mbox{Annihilation in grains}& 2.0 & 2.0 & 2.0 & 2.0 \\
\noalign{\smallskip}
  \hline
\end{array}
\end{table*}
%%%%%%%%%%%%%%%%%%%%%%%%%%%%%%%%%%%%%%%%%%%%%%%%%%%%%%%%%%%%%%%%%%%%%

However, several values of the various processes'
FWHM have now changed to some extent (by about 10 \% in most cases, but 
sometimes more) compared to the values used in previous modeling. 
In particular, the width of the line produced in the decay of Ps formed 
``in-flight" with H is now 5.8 keV instead of the 6.4 keV used for over 
two decades. (The difference is due to the accurate experimental cross 
sections now available for positron excitation, ionization and charge 
exchange with H.) Likewise the lines resulting from positron annihilation 
after capture by a grain (either two-photon annihilation inside the grain 
or decay of the Ps ejected from the grain) have widths of 2.0 keV and 
1.4 keV respectively, compared to 1.8 keV and 2.5 keV previously. 
References and discussions of the various FWHMs are given in the 
sections 3.2 and 3.3.

The following formula then allows us to construct the theoretical 
spectrum for a given ISM phase:

%%%%%%%%%%%%%%%%%%%%%%%%%%%%%%%%%%%%%%%%%%%%%%%%%%%%%%%%%%%%%%%%%%%%%%%
\begin{eqnarray}
S(E) &=& \int dE' \left[ 3 \times {3 \over 4} P_t(E') + 2 \times {1 \over 4} \delta(E'-E_0) \right] \nonumber \\
     &\{&  X \times f_{1, H/H_2} G(E, E', \Gamma_{if, H/H_2}) \nonumber \\
     &+& Y \times f_{1, He} G(E, E', \Gamma_{if, He}) \nonumber \\
     &+& \left( 1 - X \; f_{1, H/H_2} - Y \; f_{1, He} \right) \nonumber \\  
     &\times& [ f_{ce, H/H_2} \; G(E, E', \Gamma_{ce, H/H_2}) \nonumber \\  
     &+& f_{ce, He} \; G(E, E', \Gamma_{ce, He})  \nonumber \\
     &+& f_{rce} \; G(E, E', \Gamma_{rce})  \nonumber \\
     &+& f_{gr, out} \; G(E, E', \Gamma_{gr, out}) \, ]  \; \}\;  \nonumber \\
     &+& 2 \left( 1- X \; f_{1, H/H_2} - Y \; f_{1, He} \right) [ f_{dae} \; G(E, E_0, \Gamma_{dae}) \nonumber \\
     &+& f_{da, H/H_{2}} \; G(E, E_0, \Gamma_{da, H/H_2}) \nonumber \\
     &+& f_{da, He} \; G(E, E_0, \Gamma_{da, He}) \nonumber \\
     &+& f_{gr, in} \; G(E, E_0, \Gamma_{gr, in})] \; ,
    \label{eq:spec}
\end{eqnarray}
%%%%%%%%%%%%%%%%%%%%%%%%%%%%%%%%%%%%%%%%%%%%%%%%%%%%%%%%%%%%%%%%%%%%%%%

\noindent where X and Y are the relative abundances of H/H$_2$ and He (90 
\% and 10\% respectively), $ \Gamma_P $ refers to the FWHM of the line 
produced by a given process P ($if, ce, dae, daH, gr$ refer to in-flight, 
charge exchange, direct annihilation with free electrons, direct 
annihilation with H, annihilation with grains -- inside or outside --, 
etc.); $G(E, E', \Gamma_P)$ is the Gaussian function of variable E', 
centered on E, and of FWHM = $ \Gamma_P $, normalized to 1; the 
factors $f_P$ are the relative probabilities for each process: $f_P = R_P 
/ \sum R_P$, $R_P$ denoting the rates of various processes (Table 
\ref{tab:RR}) weighted by the relative abundance of each element.

%%%%%%%% Figure Y: Spectra %%%%%%%% 
%\begin{figure}[tb]
%\centering
\begin{figure*}
\includegraphics[width=8.8cm,height=7.5cm]{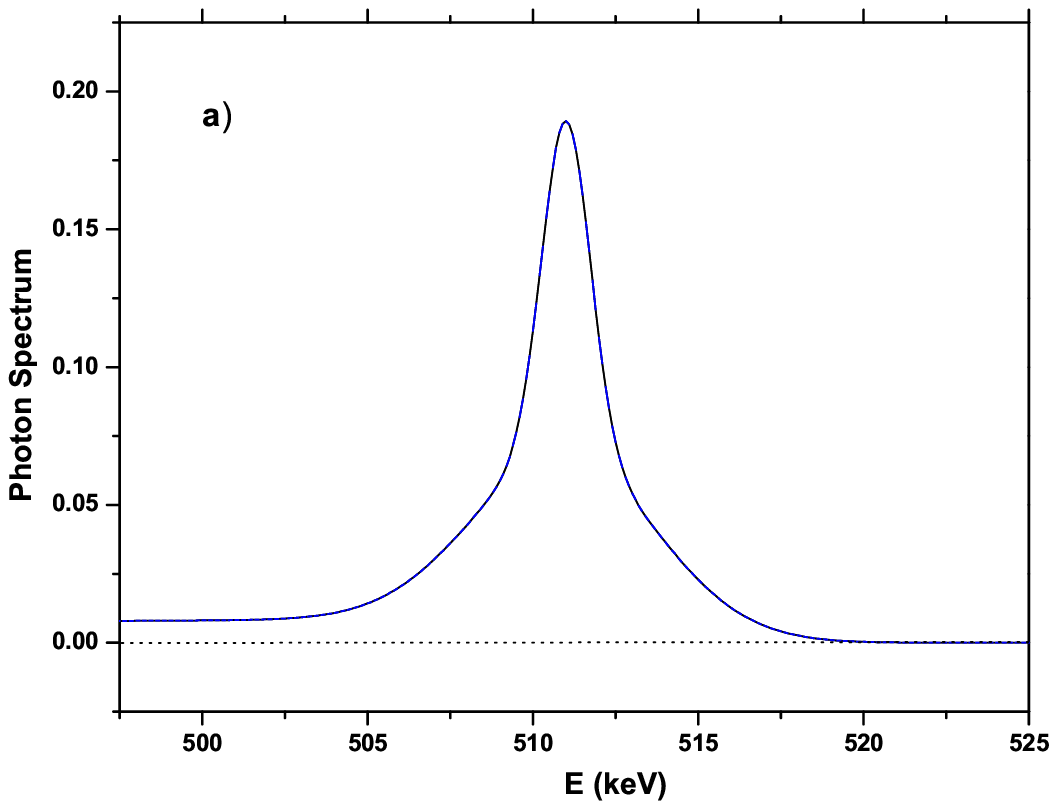}
\includegraphics[width=8.8cm,height=7.5cm]{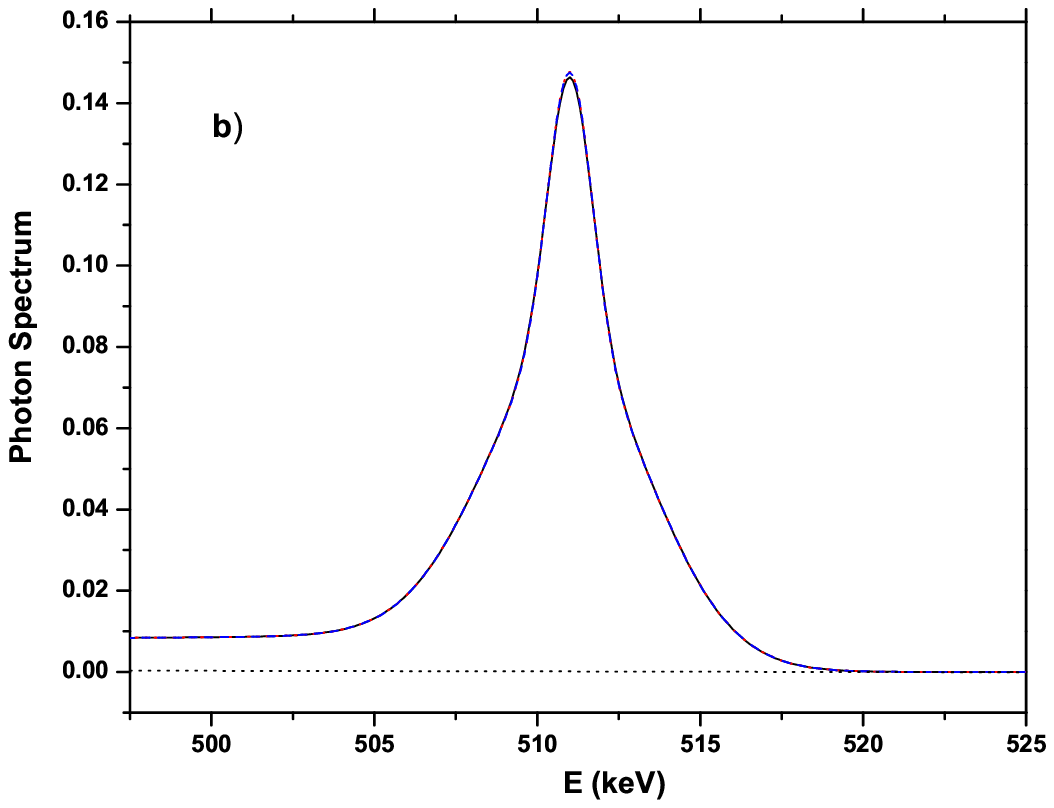}
\includegraphics[width=8.8cm,height=7.5cm]{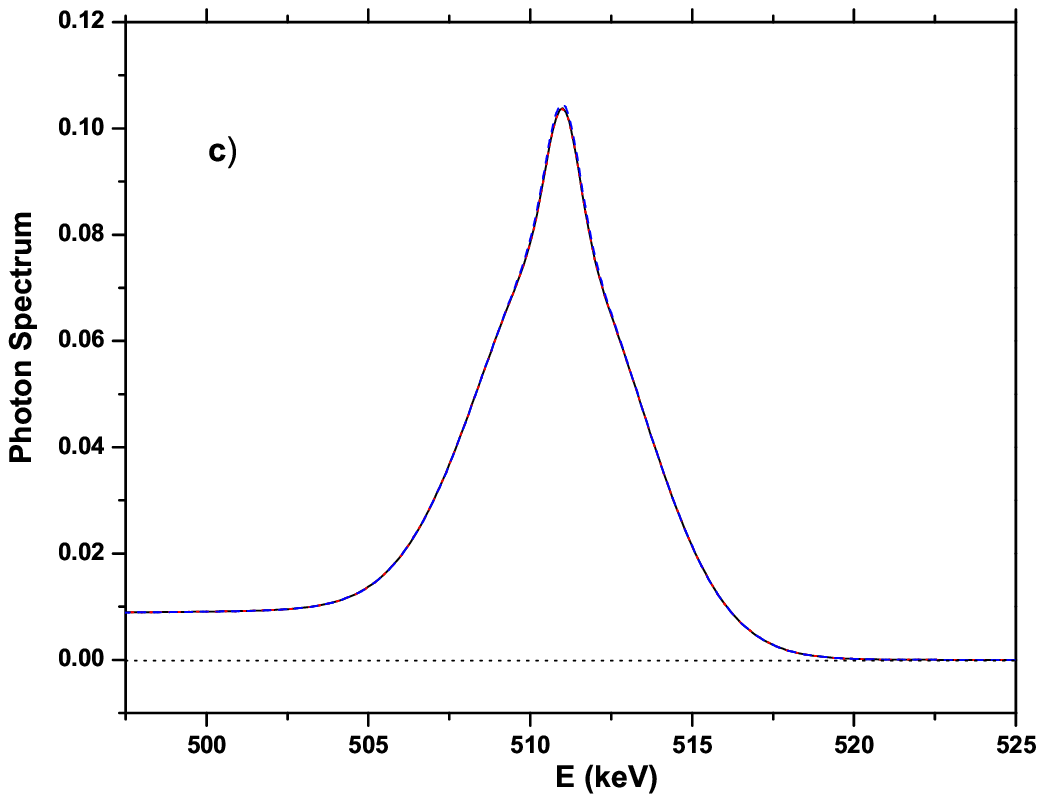}
\includegraphics[width=8.8cm,height=7.5cm]{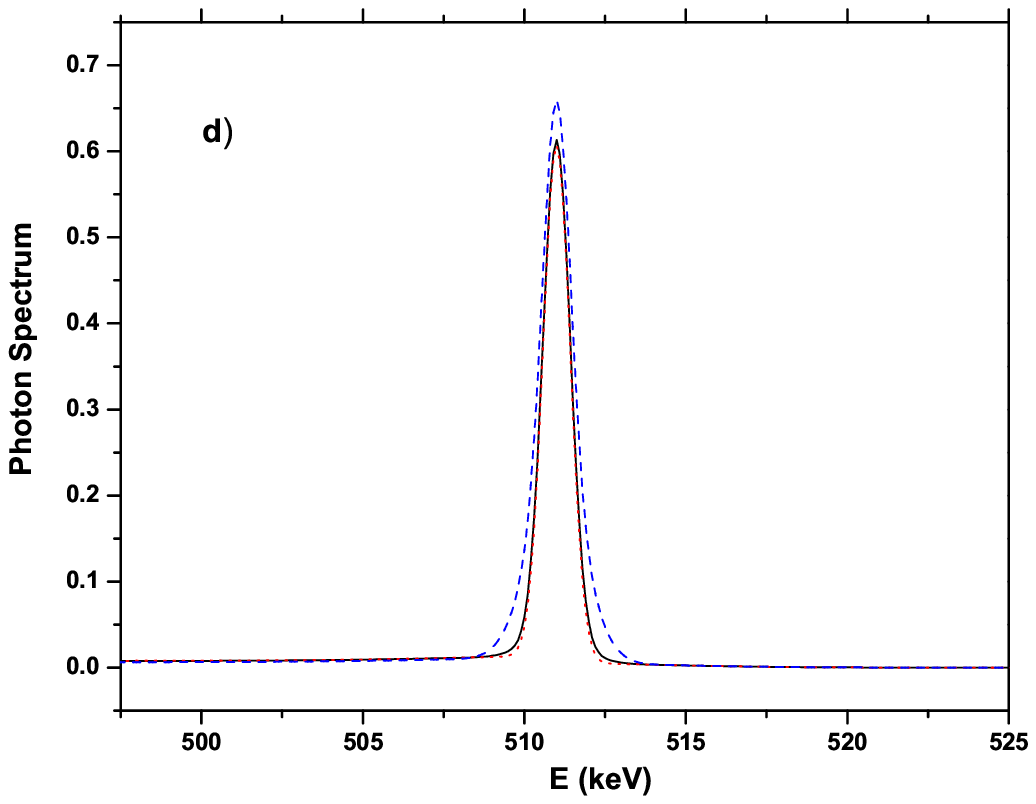}
\includegraphics[width=8.8cm,height=7.5cm]{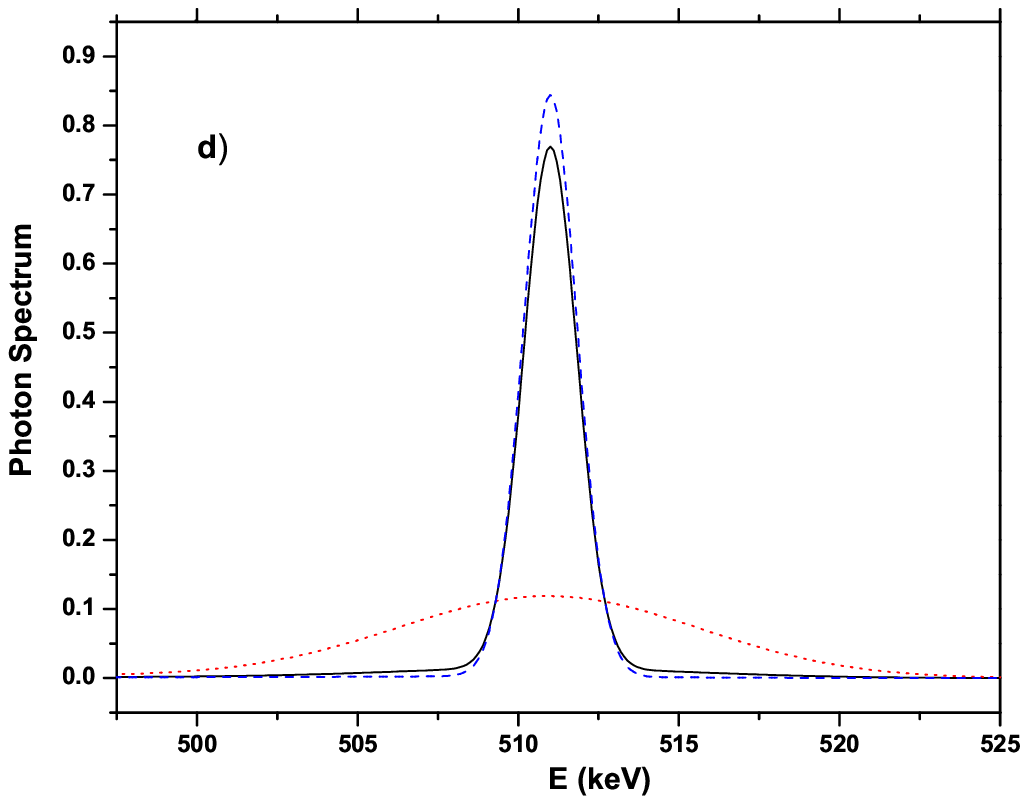}
\caption{Annihilation spectra for the 5 ISM phases separately (see text): 
(a) Cold, molecular hydrogen medium; 
(b) Cold, atomic hydrogen medium; (c) Warm, neutral medium; (d) Warm, 
ionized medium; (e) hot medium. 
\label{fig:spec}}
\end{figure*}
%%%%%%%%%%%%%%%%%%%%%%%%%%%%%%%%%%%%%%%%%%%%%%%%%%%%%%%%%%%%%%%%%%%%%%%

Figures \ref{fig:spec} (a, b, c, d, e) show the annihilation spectrum in 
the molecular, 
cold (atomic), warm neutral, warm ionized and hot phases, respectively. 
In each case the solid curve shows the spectrum assuming a ``standard 
grain model" (in terms of grain abundance, size distribution, composition, 
etc.,  as adopted in Section 3.3) in the given phase, i.e. a factor 
$x_{gr}$ = 1; dashed curves represent the spectra in the case of grains 
greatly amenable to positron capture ($x_{gr}$ = 10.0); dotted 
curves represent the spectra when dust is removed from the phase. 
The spectra confirm that unless dust is overabundant in the warm 
ionized or hot phases (or the rates of capture are greater than 
we have adopted -- as explained in section 3.3), it has a negligible 
effect on the annihilation radiation emitted in the ISM.

In order to construct the ``global" spectrum of 511 keV annihilation from 
the ISM as a whole, one simple approach is to combine the previous phase 
spectra on the basis of the relative contributions (densities and filling 
factors) of each phase. For this, filling factors such as those used in 
the ISM model of McKee \& Ostriker (1977) can be adopted: 2.4 \% for the 
cold phase(s), 23 \% for each of the warm phases and 52\% for the hot 
phase. In the same model, the densities of each phase are: 42 cm$^{-3}$, 
0.37 cm$^{-3}$, 0.25 cm$^{-3}$, $3.5 \times 10^{-3}$ cm$^{-3}$ in the 
cold, warm neutral, warm ionized and hot phases, respectively. 

\begin{figure}
\includegraphics[width=8.5cm,height=10cm]{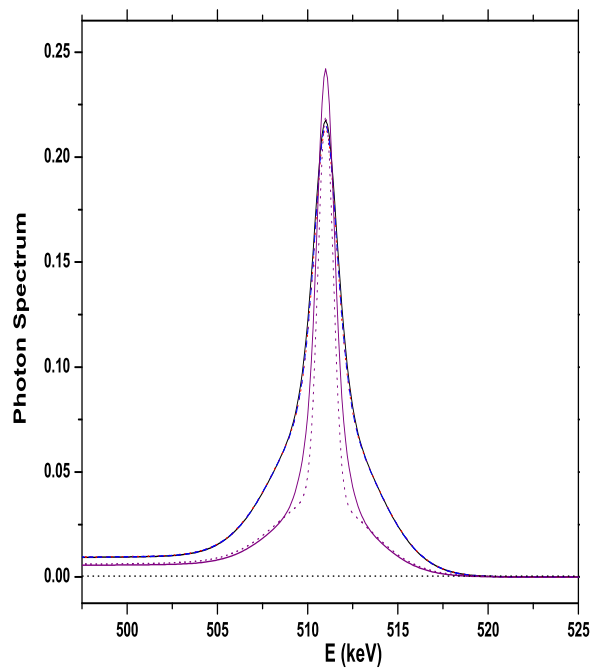}
\caption{Combined (``global") ISM annihilation spectra, obtained 
under two assumptions: 1) positrons annihilating in all phases 
according to the filling factors and densities given in the 
McKee \& Ostriker (1977) model, either with or without the effect 
of the grains (solid, blue dashed and red dotted curves); 
2) positrons not annihilating in the cold phases, with/without 
grain effect (solid and dotted narrower purple curves). 
\label{fig:speccomb}}
\end{figure}

Figure \ref{fig:speccomb} shows the result of combining the 
individual-phase spectra using the McKee \& Ostriker
simple prescription as well as the spectra obtained when 
the positrons are excluded from the cold regions (both 
the molecular and atomic phases); this we do for comparative 
purposes with the results obtained in the treatment of 
Guessoum, Ramaty and Lingenfelter (1991). In the first 
case (annihilation in all phases according to the description 
of McKee \& Ostriker), we see that the dust makes no 
difference to the line profile. In the second case, the 
profile of the overall emission spectrum is substantially 
different from the ``uniform" case, and the dust does 
make a non-negligible difference. 

%%%%%%%% Figure Z: Combined Spectrum %%%%%%%% 
%\begin{figure}[tb]
%\centering

%%%%%%%%%%%%%%%%%%%%%%%%%%%%%%%%%%%%%%%%%%%%%%%%%%%%%%%%%%%%%%%%%%%%%%%

A more physically realistic approach would be to adopt an analytic ISM 
model such as Ferriere's, which gives the densities of each species 
(H, H$_2$, free electrons, etc.) in each volume element (defined by 
its galactocentric coordinates r and z) of the ISM. This latter approach 
we save for a future, more physical treatment of annihilation in the 
ISM, one that will consider the change in the annihilation spectrum and 
profile when emanating from different directions and regions of the 
Galaxy. 

Table \ref{tab:FWHM2} displays the FWHMs of the lines obtained in each 
phase as well as in the ``global" case. These allow us to compare 
our results with earlier ones and serve to give us a rough idea 
about the physical properties of the annihilation region(s), 
although we must emphasize that the FWHM parameter is not a 
sufficient parameter in characterize such spectra. 
Until the full follow-up modeling and interpretation 
work is performed, these values are used for comparison and indication 
purposes. Our present results are largely consistent with the older 
ones, but our line widths tend to 
be slightly larger than the older ones in all phases. In the 
warm neutral phase the line appears to be substantially broader 
than the result of Guessoum, Ramaty and Lingenfelter (1991), but 
only because in that treatment the warm neutral phase has a 15 \% 
ionization fraction, whereas in the present treatment the warm 
neutral phase is completely neutral. Moreover 
the rate for charge exchange with H has increased (by about 20 \%) 
in this work, which increases the contribution of the broad line; 
indeed the in-flight process, for which the line widths are always 
large ($\approx 6$ keV), has now become more dominant overall. 

%%%%%%%% Table - Values of FWHM's obtained %%%%%%%%%%%%%%%%
\begin{table}
\caption{Values of FWHMs obtained (in keV) in each ISM phase separately 
and in the ``global" case; the values are given for the ``standard grain 
model" ($x_{gr} = 1$), for $x_{gr} = 0$ (no dust) and for $x_{gr} = 10$ 
(overabundant and/or grain characteristics especially favorable for 
positron capture -- see text), respectively. For the ``global (combined)" 
spectrum we give FWHM values in two cases: the McKee \& Ostriker 
model with its phase densities and filling fractions, and the 
situation (results in parentheses) when the positrons are excluded 
from the cold phases. \label{tab:FWHM2}}
\begin{array}[b]{lccc}
\noalign{\smallskip}
\hline
\hline
\noalign{\smallskip}
\mbox{Phase}  & \quad \mbox{$x_{gr} = 1$} & \quad \mbox{$x_{gr} = 0$} & \quad \mbox{$x_{gr} = 10$} \\
\hline
\noalign{\smallskip} 
\mbox{Molecular} & \quad \mbox{2.39} & \quad \mbox{2.39} & \quad \mbox{2.39}  \\
\mbox{Cold} &  \quad  \mbox{3.00} & \quad \mbox{3.00} & \quad \mbox{2.92} \\
\mbox{Warm Neutral} & \quad \mbox{4.78} & \quad \mbox{4.76} & \quad \mbox{4.74}  \\
\mbox{Warm Ionized} & \quad \mbox{1.02} & \quad \mbox{1.00} & \quad \mbox{1.19} \\
\mbox{Hot} & \quad \mbox{1.99} & \quad \mbox{11.0} & \quad \mbox{1.96} \\
\mbox{} &  &  \\
\mbox{Combined} & \quad \mbox{2.26 (1.18)}& \quad \mbox{2.17 (1.15)} & \quad \mbox{2.17 (1.37)} \\
\noalign{\smallskip}
  \hline
\end{array}
\end{table}
%%%%%%%%%%%%%%%%%%%%%%%%%%%%%%%%%%%%%%%%%%%%%%%%%%%%%%%%%%%%%%%

The main conclusions from this ``phase" treatment are first that the
``global" spectra obtained here (on the basis of the McKee \& Ostriker 
1977 ISM model) are consistent with the observed widths reported 
by the various missions (including SPI and TGRS) and second that unless 
specific physical assumptions are made regarding whether the positrons 
annihilate prefentially in some regions (say the hot and/or ionized phases), 
the dust will in general play a negligible role. A careful analysis and 
interpretation of the observational data in the light of our new 
modelling 
of the positron processes will be performed in a follow-up work, and we 
do not exclude a priori the possibility that positrons may show a preference 
for some phases and/or for grain capture.

%%%%%%%%%%%%%%%%%%%%%%%%%%%%%%%%%%%%%%%%%%%%%%%%%%%%%%%%%%%%%%%%%%%%%%%%%%
% 5. Summary and Conclusions
%%%%%%%%%%%%%%%%%%%%%%%%%%%%%%%%%%%%%%%%%%%%%%%%%%%%%%%%%%%%%%%%%%%%%%%%%%

\section{\label{s5}Summary and conclusions}

The aim of this work was to carefully reexamine all the processes that 
positrons undergo in the ISM, including the complicated interactions 
with dust grains, and to try to determine as precisely as possible 
all the quantitative factors in the problem, particularly the 
various reaction rates and the widths of the lines emitted in the process. 
We also included all aspects of positron interactions with molecular 
hydrogen, in addition to atomic hydrogen, as well as helium (excitation, 
ionization, charge exchange, direct annihilation with the bound electron, 
etc.). For this we undertook a major review of the literature (cross 
section measurements and calculations, etc.) and performed a detailed 
Monte Carlo simulation that allowed us to obtain new, accurate results 
on the fraction of positrons that form positronium in flight, the profile 
and width of the line emitted in the process, the lifetimes of the 
positrons in such conditions, etc. We also investigated in detail
the interstellar dust and its effect on positron annihilation. 
The final results of our calculations (reaction rates and spectra) showed 
that dust is only important in the hot phase of the ISM (where it 
dominates 
all other processes), but previous modelling efforts 
(e.g. Guessoum et al. 2004) stressed the importance of this process. 
There are still some pieces of the puzzle missing in this regard, for 
example the widths of the lines emitted in the annihilation of 
positrons/positroniums inside/outside a dust grain, the extent of 
the destruction of grains in the hot/warm regions of the ISM and 
also the electrostatic charge carried by dust in various conditions. 

We aimed to provide solid basis for modelling the detailed 
annihilation data currently being obtained by INTEGRAL-SPI 
(Knodlseder et al. 2005) as well as future observations, 
e.g. spectroscopic explorations of regions of the Galaxy other than the 
central regions (galactic plane, compact objects / LMXBs, etc.). 
Indeed, with the positron processes now much more firmly understood 
and quantified, it becomes possible to compare the annihilation model 
outlined here and the spectra obtained by our calculations  
with the SPI data and infer useful information regarding the 
regions where the annihilation is predominantly taking place. It 
will also become possible to model and interpret data obtained for 
more local and specific places. Finally, with the microscopic 
processes now fully examined and more precisely parametrized, 
it becomes possible to perform a time-dependent investigation of 
the lives of positrons from their births to their ultimate 
deaths/annihilations. This would then constitute a major step 
towards the resolution of the ``origin of positrons 
in the Galaxy" puzzle.

%%%%%%%%%%%%%%%%%%%%%%%%%%%%%%%%%%%%%%%%%%%%%%%%%%%%%%%%%%%%%%%%%%%%%%%%%%
% Acknowledgements
%%%%%%%%%%%%%%%%%%%%%%%%%%%%%%%%%%%%%%%%%%%%%%%%%%%%%%%%%%%%%%%%%%%%%%%%%%

\begin{acknowledgements}

N. Guessoum would like to acknowledge the support of the INTEGRAL project. 
The researchers and staff of the Centre d'Etude Spatiale des Rayonnements 
(Toulouse, France) are thanked for their kind hospitality during the 
various times when much of this work was conducted. Cliff Surko is thanked 
for helpful discussions and for providing us with useful references.
\end{acknowledgements}

%%%%%%%%%%%%%%%%%%%%%%%%%%%%%%%%%%%%%%%%%%%%%%%%%%%%%%%%%%%%%%%%%%%%%%%%%%

%%%%%%%%%%%%%%%%%%%%%%%%%%%%%%%%%%%%%%%%%%%%%%%%%%%%%%%%%%%%%%%%%%%%%%%%%%
% Bibliography
%%%%%%%%%%%%%%%%%%%%%%%%%%%%%%%%%%%%%%%%%%%%%%%%%%%%%%%%%%%%%%%%%%%%%%%%%%

\end{document}